\documentstyle[aps,prl,eqsecnum,preprint,tighten]{revtex}   
\begin{document}
\title{Quasiholes and fermionic zero modes of paired fractional quantum Hall
states: the mechanism for nonabelian statistics}
\author{N. Read$^1$ and E. Rezayi$^2$}
\address{$^1$ Departments of Physics and Applied Physics, 
Yale University,\\ P.O. Box 208284, New Haven, Connecticut 06520}
\address{$^2$ Department of Physics, California State University,\\ 
Los Angeles, California 90032}
\date{July 31, 1996}
\maketitle
\begin{abstract}
The quasihole states of several paired states, the Pfaffian, Haldane-Rezayi,
and 331 states, which under certain conditions may describe electrons at 
filling factor $\nu=1/2$ or 
$5/2$, are studied, analytically and numerically, in the spherical geometry, 
for the Hamiltonians for which the ground states are known exactly. We 
also find all the ground states (without quasiparticles) of these systems 
in the toroidal geometry. In each case, a complete set of linearly-independent 
functions that are energy eigenstates of zero energy is found explicitly. 
For fixed positions of the quasiholes, the number of linearly-independent
states is $2^{n-1}$ for the Pfaffian, $2^{2n-3}$ for the Haldane-Rezayi state;
these degeneracies are needed if these systems are to possess nonabelian
statistics, and they agree with predictions based on conformal field theory.
The dimensions of the spaces of states for each number of quasiholes agree 
with numerical results for moderate system sizes. The effects of tunneling and 
of the Zeeman term are discussed for the 331 and Haldane-Rezayi states, as well
as the relation to Laughlin states of electron pairs. A model introduced by Ho,
which was supposed to connect the 331 and Pfaffian states, is found to have 
the same degeneracies of zero-energy states as the 331 state, 
except at its Pfaffian point where it is much more highly degenerate than 
either the 331 or the Pfaffian. We introduce a modification of the model which
has the degeneracies of the 331 state everywhere including the Pfaffian
point; at the latter point, tunneling reduces the degeneracies
to those of the Pfaffian state.
An experimental difference is pointed out 
between the Laughlin states of electron pairs and the other paired states, 
in the current-voltage response when electrons tunnel into the edge. 
An appendix contains results for the permanent state, in which the 
zero modes can be occupied by composite bosons, rather than by composite 
fermions as in the other cases; the system is found to have an incipient
instability towards a spin-polarized state. 
\end{abstract}


\section{Introduction}
\label{introduction}
Over the past few years there has been renewed interest in 
fractional quantum Hall effect (FQHE) \cite{book} states involving 
pairing at even-denominator filling factors 
\cite{halp83,haldrez,ymg,mr,gww,wen,milr}. The earliest idea \cite{halp83} 
was to generalise the Laughlin state \cite{laugh} by first pairing
the electrons into charge-2 bosons, then forming a Laughlin state of 
the bosons, for which the filling factor $\nu_b$ must be of the form 
$\nu_b=1/m$, $m>0$ even. Since the filling factor $\nu$ of the electrons is
related to that of the bosons by $\nu=4\nu_b$ \cite{halp83}, one obtains $\nu$
either of the form $1/q$ or $2/q$, where in the second case $q$ must be odd.
For the cases $\nu=1/q$ with $q$ even, this produces a fractional quantum Hall 
state at a filling factor not accessible in the usual hierarchy theory 
\cite{hald}. This idea was taken up by Haldane and Rezayi \cite{haldrez}, 
using spin-singlet pairs, to produce a candidate to explain the observed 
$\nu=5/2$ plateau \cite{willett} (using the usual notion that filling factors 
larger than $2$ involve filling the lowest Landau level with electrons of both 
spin, and then constructing a $\nu=1/2$ state in the first excited Landau 
level). The Haldane-Rezayi (HR) wave function has a simple structure, and other
paired states with analogous structures, for either spin-singlet or
spin-polarized pairs, were written down in Refs.\ \cite{ymg,mr}. In particular,
the Pfaffian state of Moore and Read \cite{mr} is the simplest paired state for
spinless or spin-polarized particles. The latter authors also argued that the 
paired states exhibit pairing of composite particles, either bosons or 
fermions, constructed by attaching an odd or even number $q$, respectively, 
of vortices to the electrons, for filling factor $1/q$ \cite{read,jain}. 
These objects behave like particles in zero magnetic field,
and the wave functions of the paired states can be interpreted as
Bardeen-Cooper-Schrieffer (BCS) paired wave functions, in their position space
form. In particular, this makes it easy to understand why the HR state is a
spin-singlet. It was also suggested that there should be low-energy excitations 
in which composite particles are unpaired (but still consist of electrons 
attached to vortices), as
opposed to breaking the electron pairs in Halperin's picture. It was further
suggested that quasiparticle excitations analogous to those of the Laughlin
state \cite{laugh}, which in incompressible states correspond to vortices in
the order parameter \cite{read}, would carry multiples of a half flux quantum,
and thus charges in multiples of $1/2q$, rather than $1/q$ as in the Laughlin
states at filling factor $1/q$ (this also results {}from viewing the excitations
as quasiparticles in the Laughlin states of charge-2 bosons).
Finally, it was proposed \cite{mr} that these quasiparticles 
obey nonabelian statistics. In brief, nonabelian statistics requires that
there be degenerate states for well-separated quasiparticles, and when the
quasiparticles are exchanged adiabatically, the effect is not merely a Berry
phase representing ordinary fractional statistics, but a matrix acting within
the space of degenerate quasiparticle states. In the present paper, we do not
aim to exhibit this action directly, but we do aim to show that the quasihole
states of the Pfaffian and HR states possess the necessary degeneracies, and 
to give a physical explanation of their origin.

In subsequent work by Greiter, Wen and Wilczek \cite{gww}, the physics of 
the formation of the Halperin-type paired states, that is Laughlin states 
of electron pairs, was elaborated, using the Moore-Read Pfaffian state 
as an example, 
and several of the points made in Ref.\ \cite{mr} were repeated.
Greiter et al.\ also introduced a three-body Hamiltonian for which the Pfaffian
state at $\nu=1$ is the exact zero-energy eigenstate.
As regards the statistics of the quasiparticles, however, Greiter et al.\
argued, quite reasonably, that Halperin's picture would lead to simple abelian
statistics of the quasiparticles. While we agree with much
of the physical discussion by these authors (including the argument that the
Halperin paired states will have $4q$-fold-degenerate ground states on the 
torus in the thermodynamic limit), we disagree with their use of the three-body 
Hamiltonian and Pfaffian-based simple wave functions to illustrate their 
points. Other work on this model \cite{wen,milr}, and even the observation 
by Greiter et al.\ themselves \cite{gww} that there is a six-fold degeneracy 
of zero-energy states of the three-body interaction on the torus, are more 
consistent with the predictions of Ref.\ \cite{mr} of nonabelian statistics 
and related properties that are connected with conformal field theory (CFT) 
in two spacetime dimensions. For example, there are gapless Majorana fermion 
excitations at an edge of the Pfaffian state \cite{wen,milr}, in addition to 
the usual charge-fluctuation boson excitations, while the Halperin-type state 
of electron pairs would be expected to possess only the latter. The results 
that will be obtained in the present paper lend further support to the belief
that the quasiparticle states that are constructed as energy eigenstates of the 
three-body Hamiltonian of Ref.\ \cite{gww} (and its generalizations to be
constructed below) do possess nonabelian statistics. 

It was also suggested \cite{gww} that the nonabelian behavior might be 
present only at points of special symmetry, and not be generic. Clearly, the
three-body Hamiltonian might be such a point. Although it was argued in Ref.\ 
\cite{mr} that nonabelian statistics is a topological property that cannot be
altered by small perturbations because the ground states involved are assumed
to have an energy gap for all excitations, this has not been tested. It is
clearly an important problem, but it lies beyond the scope of the present
paper.     

A further development in paired FQHE states was the realization that some
members (to be referred to here collectively as
the 331 state) of another class of states, of which an example was 
introduced by Halperin \cite{halp83}, also exhibit pairing 
\cite{haldrez,gww2,halpnewport}. These states have
come under scrutiny because of their relevance for FQHE states in double-layer
systems at $\nu=1/2$ \cite{hr87,ymg2,gww2,he}. They can also be viewed as 
generalized hierarchy states
\cite{read90}, and so are not expected to possess nonabelian statistics,
however, they are still distinct {}from the Halperin idea of a Laughlin state of
charge-2 bosons. We will discuss these states, and especially recent work by
Ho \cite{ho}, further in Sec.\ \ref{tunneling}.

As we have mentioned above, the main purpose of this paper is to check 
the expectation, based on the CFT ideas of Ref.\ \cite{mr}, that the quasihole
states of the Pfaffian and HR states possess degeneracies above and beyond
those that would be obtained for ordinary Laughlin quasiholes, or their
generalization to the Halperin-type paired states, and thus to lay the
groundwork for a demonstration of nonabelian statistics. We do this by
constructing zero-energy eigenstates of those Hamiltonians for which the simple
form of ground state wave function is correct. These systems serve as model
examples, each of which we may hope is typical of a universality class (in the
sense of ref.\ \cite{mr}), though the study of the effect of perturbations lies
beyond the scope of this paper. 
(We should mention that quasielectrons are expected to obey nonabelian
statistics like those for the quasiholes, but it is always much more difficult
to obtain energy eigenfunctions, of which the wave functions take 
a nice form, for quasielectrons than for quasiholes, and the energies will not
be zero, nor degenerate, though presumably the degeneracies would be recovered 
in the thermodynamic limit for well-separated quasielectrons.) 
Some of the results for the Pfaffian appeared
in an unpublished earlier work \cite{rrunpub} (see also Ref.\ \cite{nayak}), 
but the method employed here in
general is related to that used for the edge states in Ref.\ \cite{milr}.
The results are in full accord with the earlier expectations. It will emerge
that the degeneracies of quasihole states of the paired states can be viewed as
coming {}from breaking pairs of composite particles and placing the unpaired
(composite) particles in certain single-particle states that contribute 
zero to the total energy; these are ``zero modes''. 

Throughout this paper we will use the terminology ``particles'' to refer to the
underlying charged particles in the lowest Landau level, which could be either
fermions (such as electrons) or bosons, and not to the composite particles. 
For a given Hamiltonian, we will also refer to energy eigenstates that have 
energy eigenvalue equal to zero simply as zero-energy states.

In this paper we mostly work on the sphere; we will briefly review this
formalism, and results for the quasiholes of the Laughlin state \cite{hald}. 
One uses a uniform radial magnetic field with a total of $N_\phi$ flux 
through the surface, and in the lowest Landau level (LLL) each particle has 
orbital angular momentum $N_\phi/2$. The LLL wave functions on a sphere 
are usually written (in a certain gauge \cite{hald}) in terms of ``spinor'' (or
``homogeneous'') 
coordinates $u_i$ and $v_i$ for each particle $i=1$, \ldots, $N$, with 
$u_i=e^{i\phi_i/2}\cos\theta_i/2$, $v_i=e^{-i\phi_i/2}\sin\theta_i/2 $ in 
terms of the spherical polar coordinates $\theta_i$, $\phi_i$, on the sphere. 
Since these imply that $u_i$, $v_i$ are not independent complex numbers, 
it is often more convenient, and will simplify the writing, to use a 
nonredundant parametrization of the sphere by a single complex variable. 
This is done by stereographic projection, which gives the definition 
$z_i=2Rv_i/u_i$, where $R$ is the radius of the sphere. Single-particle 
basis states in the LLL then take the form 
$z_i^m/(1+|z_i|^2/4R^2)^{1+N_\phi/2}$, where the $L_z$ angular momentum 
quantum number is $L_z=N_\phi/2-m$. In this form, the rotationally-invariant 
inner product of single-particle states on the sphere is given by multiplying 
one function by the conjugate of the other and integrating over the $z_i$ plane 
with no other $z_i$-dependent factors inside the integral. Only 
single-particle basis states with $m\leq N_\phi$ correspond to LLL functions 
on the sphere and can be normalized with respect to this inner product
(the normalizing factors will not be needed here). [Note that when $N_\phi$ and
$R\rightarrow\infty$ with $N_\phi/R^2$ fixed, in which case the sphere becomes
effectively flat, the basis functions (for $m$ fixed) tend to 
$z_i^m e^{-|z_i|^2/4}$, the basis functions in the plane in the symmetric 
gauge \cite{laugh}.] Many-particle states can thus be
written as 
\begin{equation}
\Psi=\tilde{\Psi}\prod_i(1+|z_i|^2/4R^2)^{-(1+N_\phi/2)}
\end{equation}
and $\tilde{\Psi}$ must be a polynomial of degree no higher than $N_\phi$ in
each $z_i$. Therefore, in the following we need specify only $\tilde{\Psi}$ in
order to describe a state.

The Laughlin ground state and quasihole states are exact, zero-energy states
for short-range pseudopotential Hamiltonians \cite{hald} of the general form
\begin{equation}
H=\sum_{i<j}\sum_{M=0}^{N_\phi} V_M P_{ij}(N_\phi-M)
\label{pseudH}
\end{equation}
in which $P_{ij}(L)$ is a projection operator onto the subspace in which the
total orbital angular momentum of the particles $i$ and $j$ is $L$; in the
summation, $M$ can be viewed as the relative angular momentum. For the LLL
states, close approach of the two or more particles occurs only when their 
total angular momentum is large. The parameters $V_M$ are the pseudopotentials.
Note that for spinless particles, only even $M$ occur for bosons, and only odd
$M$ for fermions. We will later generalise the projection operator notation 
to three-body operators $P_{ijk}$, and the subspace onto which it projects 
will be specified by the values of further quantum numbers of the chosen group 
of particles, such as total spin $S$, or specific values 
of the $z$ component of the spin of each particle, etc. Every projection
operator is always normalized in the conventional way, with $P^2=P$.

The Laughlin states are zero-energy states for the pseudopotential Hamiltonian 
in which $V_M\neq 0$ for $M<q$ and zero otherwise (in fact, the non-zero $V_M$
are usually taken to be positive). The
Laughlin-Jastrow wave function is
\begin{equation}
\tilde{\Psi}_{\rm LJ}=\prod_{i<j}(z_i-z_j)^q.
\end{equation}
Clearly $q$ must be even when the particles are bosons, odd when they 
are fermions. The number of flux is then $N_\phi=q(N-1)$, and
the filling factor, $\nu=N/N_\phi$, tends to $1/q$ as the number of
particles $N\rightarrow \infty$. We will always use the integer $q>0$ as the 
parameter specifying the filling factor $\nu=1/q$. In this state, any 
two particles have relative angular momentum $M\geq q$ \cite{hald}, so it is
annihilated by $H$. This property is preserved if the state is multiplied by
the quasihole factors $U(w)=\prod_i(z_i-w)$, which change the flux by one
quantum. These factors can be expanded in powers of each $w$ to obtain 
the elementary symmetric polynomials in the $z_i$'s,
\begin{equation}
e_m=\sum_{i_1<i_2<\cdots<i_m}z_{i_1}z_{i_2}\ldots z_{i_m}
\end{equation}
which are linearly-independent operators, 
and the states obtained by
multiplying in several of these factors span the full space of zero-energy
states for each number of flux $N_\phi=q(N-1)+n$, where $n$ is the number of
quasiholes \cite{laugh,hald}. This results {}from the standard fact about
symmetric polynomials that they can all be obtained as sums of
products of the elementary symmetric polynomials. 
The space of states obtained in this way is equivalent to that for $N$ bosons
on the sphere in the lowest Landau level with $n$ flux, or $n+1$ orbitals. This
can be viewed as the $q=0$ case of Laughlin's states, which applies since the
dimension of the desired space of states is independent of $q$. The dimension
of the space is therefore a binomial coefficient
\begin{equation}
\left(\begin{array}{c}
N+n\\
n
\end{array}\right).
\label{binom}
\end{equation}
Also, the expansion of
\begin{equation}
\prod_i\prod_k(z_i-w_k)
\label{manyqholes}
\end{equation}
in sums of products of symmetric polynomials in the $w$'s shows that, when the 
$w$'s are viewed as the coordinates of bosons \cite{hald}, the space of 
available states for these bosons, which behave as if in their LLL with $N$ as 
the number of flux, exactly coincides with the space of zero-energy quasihole 
states. The dimension of this space is then given by the formula for $n$ 
bosons in $N+1$ orbitals, which is the same binomial coefficient (\ref{binom}).
The equivalence of these viewpoints is the basic duality between bosonic 
particles and vortices within the LLL; it is analogous to the 
particle-hole transformation for fermions. The count can also be
performed by using the $q=1$ (fermion) case instead of the $q=0$ (boson) case. 
It then gives the number of states for $N$ fermions in $N_\phi+1=N+n$ 
orbitals, or for $n$ holes obeying Fermi statistics in the same
number of orbitals, and these are the same number (\ref{binom}).
We will often refer to the dimension of the spaces of zero-energy states we
find in this paper simply as the number of zero-energy states.

We now summarise the contents of the remainder of this paper. In Secs.\
\ref{pfaffian}, \ref{HR}, and \ref{331} we study the quasihole states of the
Pfaffian, HR, and 331 states on the sphere, for the Hamiltonians for which
these ground states are exact. We find explicit wave functions for 
all the quasihole states, and count them to exhibit in particular the
degeneracy that occurs even when the positions of the quasiholes are fixed. 
For the Pfaffian and HR states, this is related to nonabelian statistics, while
for the 331 states it results simply {}from a layer quantum number of the
quasiholes. The analytical results are confirmed numerically. 
In Sec.\ \ref{torus}, we consider the ground states of the same
Hamiltonians on the torus, that is with periodic boundary conditions, and
obtain the wavefunctions of the zero energy states in all cases. In Sec.\
\ref{tunneling}, we make a modest attempt to discuss the effects of 
perturbations on the states considered, especially the Zeeman term 
(for systems like HR that include particles of both spin) and tunneling 
(for double layer systems like 331). We make a full analysis of the 
zero-energy states of a model proposed by
Ho \cite{ho}, which we show to be compressible and thus pathological at the 
point where the spin-polarized Pfaffian is among the zero-energy states. 
We also rectify this problem by adding further terms to the Hamiltonian. 
These results are again checked numerically. In addition, we mention an 
experimental test that can distinguish the Halperin and other paired states, 
by using electron tunneling into the edge, for example via a point contact.
Appendix \ref{theta} contains definitions used in Sec.\ \ref{torus}, and 
Appendix \ref{perm} analyses a further paired state, the permanent state, 
in which there are spin-singlet pairs of spin-1/2 composite bosons \cite{mr}; 
this state is found to be at a transition point to a ferromagnetically-ordered 
state.

\section{Quasiholes of the Pfaffian state on the sphere}
\label{pfaffian}

In this section we will obtain all the energy eigenstates at zero energy for
the three-body Hamiltonian for which the Pfaffian ground state is exact, for
arbitrary numbers of added flux, that is for any number of quasiholes. The
following sections generalize the results to the Haldane-Rezayi and 331 states, 
and (partially) to the torus.

The Pfaffian state \cite{mr}, for even particle number $N$, is defined by the
wave function           
\begin{equation}
\tilde{\Psi}_{\rm Pf}(z_{1},\cdots,z_{N})=
{\rm Pf}\,\left(\frac{1}{z_{i}-z_{j}}\right)
\prod_{i<j}(z_{i}-z_{j})^{q},
\end{equation}
where the Pfaffian is defined by
\[{\rm Pf}\, M_{ij}=\frac{1}{2^{N/2}(N/2)!}
\sum_{\sigma\in S_{N}} {\rm sgn}\, \sigma \prod_{k=1}^{N/2} M_{\sigma(2k-1) 
\sigma(2k)}\]
for an $N\times N$ antisymmetric matrix whose elements are $M_{ij}$; $S_N$ is
the group of permutations of $N$ objects. The filling factor is $1/q$. 
The Pfaffian state is totally antisymmetric 
for $q$ even, so could describe electrons, while for $q$ odd it describes 
charged bosons in a high magnetic field. For $q=1$, it 
is the zero-energy state of the lowest flux of the Hamiltonian \cite{gww}
\begin{equation}
H=V \sum_{i<j<k}\delta^{2}(z_{i}-z_{j})\delta^{2}(z_{i}-z_{k}),
\label{pfaff3bodH}
\end{equation}
where the sum is over distinct triples of particles. 

For numerical purposes on the sphere, it is more convenient to work with a 
projection operator form of the three-body Hamiltonian, instead of 
the $\delta$-functions in (\ref{pfaff3bodH}). The closest
approach of three particles on the sphere corresponds to the state of maximum
possible total angular momentum for the three. If the particles are bosons, 
the largest possible total angular momentum is $3N_\phi/2$ (recall that each
particle has angular momentum $N_\phi/2$). Then, for the $q=1$ case, the 
Hamiltonian may be taken as proportional to the projection operator onto the 
(unique) multiplet of maximum angular momentum for each triple of bosons:
\begin{equation}
H=\sum_{i<j<k}V P_{ijk}(3N_\phi/2).
\label{pfaff3bodprojH}
\end{equation} 
The same trick works for the three-body interaction of fermions giving 
the $q=2$ case; in this case, the maximum total angular momentum of three
particles is $3N_\phi/2-3$. Some numerical results for such Hamiltonians were 
already given in \cite{wen}. For larger $q$, these Hamiltonians
can be generalized, in such a way that the zero-energy states are obtained 
{}from
those for $q=1$ by multiplying by $\prod(z_i-z_j)^{q-1}$ (it is assumed that
for $q$ odd, we are discussing bosons, and for $q$ even, fermions). The 
presence of the latter factor implies that they are all zero-energy 
eigenstates of the projection operators for any two particles onto relative 
angular momentum $M=0$, $2$, \ldots, $q-3$ ($q$ odd), or $M=1$, $3$, \ldots, 
$q-3$ ($q$ even) [or the corresponding total angular momenta $N_\phi$, 
$N_\phi-2$, \ldots, $N_\phi-q+3$, ($q$ odd), etc.]. The space of states 
annihilated by such projections is in one-to-one correspondence with the full 
space of states of the $q=1$ case, and the desired three-body projection 
operator [onto angular momentum $3N_\phi/2-3(q-1)$] is the unique one that
corresponds under this mapping to that already mentioned for $q=1$. For each
$q$, the Hamiltonian can then be taken to be the sum of the three-body and 
all of these two-body projection operators. A very similar approach works for 
the other Hamiltonians studied in this paper, so that results for higher $q$ 
can be deduced easily {}from those for the minimal $q$.
These Hamiltonians can also be written in terms of delta functions and their
derivatives, so as to arrive at a form suitable for use in geometries other 
than the sphere.
An attempt at a Hamiltonian appropriate for the Pfaffian at $q=2$ in 
the second paper in Ref.\ \cite{gww} is invalid as it annihilates all states.

In Fig.\ \ref{fig:pfg(r)} we show the two-particle correlation function 
$g(r)$ for the Pfaffian
state on the sphere with $q=2$ for three sizes, $N=10$, $12$, $14$. 
We have plotted the function versus the great circle separation $r$ (in units
of magnetic length) on the sphere, so that the largest possible value of $r$ 
is half the circumference, and we have normalized the curves in such a way 
that in an infinite system they would approach $1$ at infinity. We see that, 
although for $N=10$ an exponential decay at large distances is not apparent, 
for $N=12$ and $14$ the curve appears to be rapidly approching 1 at large 
separation, and these two curves are almost indistinguishable, in the region 
where both are defined. The correlation length in the Pfaffian, which would be
defined as the length over which $g(r)-1$ decays by a factor of $e$, 
is apparently quite large. 

The Pfaffian state is the only zero-energy eigenstate of $H$ at
$N_\phi=q(N-1)-1$. Zero-energy quasihole excitations can be obtained
only by increasing $N_\phi$, as for the quasiholes of the Laughlin state as
discussed in Sec.\ I, but in this case the basic objects contain
a half flux quantum each and must be created in pairs. A wave function for two
quasiholes was proposed in \cite{mr}; it is 
\begin{eqnarray}
\lefteqn{\tilde{\Psi}_{\rm Pf\, +\, 2\, qholes}(z_{1},\ldots,z_{N};w_{1},w_{2})
=} \nonumber \\
 & & \frac{1}{2^{N/2}(N/2)!}\sum_{\sigma\in S_{N}} {\rm sgn}\, \sigma 
\frac
{\prod_{k=1}^{N/2} [(z_{\sigma(2k-1)}-w_{1})(z_{\sigma(2k)}-w_{2})+(w_{1}
\leftrightarrow w_{2})] }
{ (z_{\sigma(1)}-z_{\sigma(2)})\cdots(z_{\sigma(N-1)}-z_{\sigma(N)}) }
 \nonumber \\
 & & \qquad\times\prod_{i<j}(z_i-z_j)^q.
\label{pfaff2qholes}
\end{eqnarray}              
It is clearly the pairing structure built into the ground state
which allows insertion of Laughlin-like factors 
\begin{equation}
f(z_i,z_j;w_1,w_2)=(z_i-w_1)(z_j-w_2)+(z_i-w_2)(z_j-w_1)
\label{2qholes}
\end{equation}
which act only on one member of
each pair, and, since the $f$'s increase the maximum angular momentum for each
$z_i$ by 1, $N_\phi$ increases by 1. As the quasiholes are, at least
approximately, located at $w_1$, $w_2$, they effectively contain a half flux 
quantum each, unlike the usual Laughlin quasihole that corresponds to a full 
flux quantum. The same structure requires that quasiholes are made in pairs, 
since the wave function must be totally symmetric or antisymmetric. 
When quasiholes coincide, that is, when $w_1=w_2$, a Laughlin quasihole is
recovered. 

It is clear that by inserting more factors $f$, with different $w$'s, into the
sum over permutations, a whole host of zero-energy eigenfunctions can be
obtained. However, this involves dividing the quasihole coordinates $w_1$,
\ldots, $w_{2n}$ into pairs in an arbitrary way; the resulting functions are
invariant only under exchanges of the two quasihole coordinates in each of these
pairs, or under permutatations of the pairs. One must then ask whether all
these states, of which there are $(2n)!/(2^n\,n!)=
(2n-1)(2n-3)\cdots\equiv(2n-1)!!$, are linearly 
independent, and also whether all zero-energy eigenfunctions can be obtained 
in this way. 

For four quasiholes, the three distinct functions obtained {}from dividing
the $w$'s into pairs in three distinct ways obey just one linear relation, as we
will now show. (These methods and results for four quasiholes appeared
previously in Ref.\ \cite{rrunpub}.) For more than four quasiholes, the 
following method becomes
increasingly impractical, and we will instead use a more direct method, inspired
by the results for edge states of the Pfaffian in \cite{milr}.

It is convenient to write the functions in the more general form
\begin{equation}
\tilde{\Psi}_p(z_1,\ldots,z_N;w_1,\ldots,w_{2n})
=\hbox{Pf}\,\{\Phi_p(z_i,z_j;w_1,\ldots,w_{2n})/(z_i-z_j)\}
                     \tilde{\Psi}_{\rm LJ}.
\label{genfunc}
\end{equation}
Here the $\Phi_p$ must be symmetric and of degree $n$ in 
$z_1$, $z_2$ in order to represent $2n$ quasiholes,
that is so that $N_\phi=q(N-1)-1+n$.
We could use products of the $f$'s in eq (\ref{2qholes}), but
it is convenient to use the following (these choices are clearly related by
taking linear combinations). For $n=2$, define 
\begin{eqnarray}
\Phi_1(z_1,z_2;w_1,\dots,w_4)&=&(z_1-w_1)(z_1-w_2)(z_2-w_3)(z_2-w_4)
\nonumber\\  
                          & &\mbox{}+(z_1-w_3)(z_1-w_4)(z_2-w_1)(z_2-w_2),    \\
\Phi_2(z_1,z_2;w_1,\dots,w_4)&=&\Phi_1(z_1,z_2;w_1,w_3,w_2,w_4),             \\
\Phi_3(z_1,z_2;w_1,\dots,w_4)&=&\Phi_1(z_1,z_2;w_1,w_4,w_2,w_3).             
\end{eqnarray}
The following identity is useful: For any set of complex numbers
$a_i$, $i=1,\ldots N$, $N$ even, $>2$,
\begin{equation}
\hbox{Pf}\,(a_i-a_j)=0,
\label{pfaffident}
\end{equation}
since the square of the Pfaffian is a determinant in which any three rows 
or columns obey a linear relation. 
Set $\Phi=\Phi_1+\Phi_2+\Phi_3$, then using (\ref{pfaffident}) it 
can be shown that
\begin{equation}
\Phi'_1\equiv \Phi_1-\frac{1}{3}\Phi={\frac{1}{3}}(z_1-z_2)^2[(w_1-w_4)(w_2-w_3)
                       +(w_1-w_3)(w_2-w_4)].
\label{Fident}
\end{equation}
Hence $\Phi_2-\Phi_3=x(\Phi_1-\Phi_2)$, where 
\begin{equation}
x=(w_1-w_2)(w_3-w_4)/(w_1-w_4)(w_2-w_3)
\end{equation}
is the cross-ratio. Thus as functions of $z_1$ and $z_2$, $\Phi_1$, $\Phi_2$, 
$\Phi_3$ 
are linearly related.
To show there are no further relations, consider the limit $w_1\rightarrow
w_2$, $w_3\rightarrow w_4$. We find
\begin{eqnarray}
\Phi_1&\sim&(z_1-w_1)^2(z_2-w_3)^2+(z_1-w_3)^2(z_2-w_1)^2,\nonumber\\
\Phi_2\sim \Phi_3&\sim&2(z_1-w_1)(z_1-w_3)(z_2-w_1)(z_2-w_3),
\end{eqnarray}
which are clearly linearly independent.

For $N=2$ particles, it now follows immediately that there are only two 
linearly independent states of the type shown. For arbitrary even $N$, we 
still have to prove that
$\hbox{Pf}\,\{\Phi_p(z_i,z_j;w_1,w_2,w_3,w_4)/(z_i-z_j)\}$ gives only two 
linearly independent states for fixed $w$'s. For any $p=1$, $2$, $3$, we use 
(\ref{Fident}) and expand the Pfaffian in powers of $\Phi'_p$. All terms 
containing more 
than one factor of $\Phi'_p$ cancel using the identity (\ref{pfaffident}) 
since they contain factors $(z_i-z_j)$. Therefore {\em the 
$N$-particle wave functions satisfy the same linear relation as the $\Phi_p$},
for all $N$. A similar argument shows that use of linear combinations of the 
$\Phi_p$ inside the Pfaffian leads only to linear combinations of the same 
states. We note that the linearly-independent states can be taken to be the 
unique state where $\Phi_p$ in eq.\ (\ref{genfunc}) is replaced by $\Phi$, and 
that where only one factor of $\Phi_p$ in the expansion of the wave function is 
replaced by $\Phi_1'$, the other $\Phi_p$ being replaced by $\Phi$. The effect 
of $\Phi_1'$ is to cancel the pairing factor $(z_{\sigma(2k-1)}-
z_{\sigma(2k)})^{-1}$ for the pair on which it acts. Thus there is a 
``broken pair'' in the wave function, as in \cite{milr}. This observation 
provides the method to generalise these results to any number of quasiholes. 
As the $w_i$ vary, these states span a space of zero-energy 4-quasihole states,
whose dimension we will find below when we have results for general $n$.

We now turn to the method for arbitrary numbers of quasiholes or added flux
quanta. We will first write down the functions, then explain why they both span
the full vector space of zero-energy eigenstates for each $n$ and $N$, and are
linearly independent. The functional form was inspired by those in \cite{milr}.
The functions are defined as
\begin{eqnarray}
\lefteqn{\tilde{\Psi}_{m_1,\ldots,m_F}(z_1,\ldots,z_N;w_1,\ldots,w_{2n})=}
\nonumber\\
& & \frac{1}{2^{(N-F)/2}(N-F)/2!}\sum_{\sigma\in S_{N}}{\rm sgn}\,\sigma 
\prod_{k=1}^F
z_{\sigma(k)}^{m_k}\prod_{\ell=1}^{(N-F)/2} \frac{\Phi(z_{\sigma(F+2\ell-1)},
z_{\sigma(F+2\ell)};w_1,\ldots,w_{2n})}
{(z_{\sigma(F+2\ell-1)}-z_{\sigma(F+2\ell)})}
\nonumber \\
 & & \qquad\times\prod_{i<j}(z_i-z_j)^q.
\label{pf+2nqholes}
\end{eqnarray}              
In this equation, $\Phi(z_1,z_2;w_1,\ldots,w_{2n})$ is defined so as to be
symmetric in the $w$'s, and is a generalization of the function $\Phi$ used in
the $n=2$ case:
\begin{equation}
\Phi(z_1,z_2;w_1,\ldots,w_{2n})=
\frac{1}{(n!)^2}
\sum_{\tau\in S_{2n}}\prod_{r=1}^n(z_1-w_{\tau(2r-1)})(z_2-w_{\tau(2r)}).
\label{Phidef}
\end{equation}
Clearly the integers $m_k$ must obey $0\leq m_k \leq n-1$ for each
$k=1$, \ldots, $F$, since the flux is $N_\phi=q(N-1)-1+n$, and can be taken to 
be ordered and distinct, $0 \leq m_1<m_2<\cdots<m_F\leq n-1$, because of the
antisymmetrization by the sum over permutations; thus $0\leq F\leq n$. Clearly
we must also have $F\leq N$; for $N\geq n$ this restriction can be ignored, 
and the analytic formulas below apply in this limit. 
A similar caveat applies to the other paired states below, but will
not be explicitly mentioned after this section.
These functions (\ref{pf+2nqholes}) represent pairing but with $F$ fermions 
left unpaired.
One could think of the unpaired fermions as resulting {}from breaking pairs 
for $N$ even, but the states also make sense for $N$ odd (note that $N-F$ is 
always even). These functions are closely analogous to the excited
quasiparticle states of a BCS paired system, where the unpaired particles
usually occupy plane waves. A whole spectrum of such excitations is expected
also in the paired FQHE states \cite{mr,gww} for any number of quasiholes
(including zero), but these generally have non-zero energy. Here we are
interested only in the subset that have zero energy for the three-body
Hamiltonian, which occur only when quasiholes are present. These states contain
fermions occupying a certain set of single-fermion wave functions $1$, $z$,
$z^2$, \ldots, $z^{n-1}$, which can be viewed as the LLL wave functions
for a flux of $n-1$, or as an angular momentum multiplet of angular momentum
$(n-1)/2$. However, the actual spatial distribution of the unpaired 
fermions in these states is hard to calculate since it must take into account 
the whole many-particle wave function. Since the effective magnetic field seen
by the fermions is essentially zero except in the quasiholes where the density
is lower, we expect that the orbitals have weight concentrated on the
quasiholes. The occupation of these orbitals 
contributes nothing to the energy for our Hamiltonian, so these are 
``zero modes''.  The number of zero modes is $n$, the number of added (real) 
flux quanta, and not $2n$, the number of quasiholes. Thus one cannot say that 
there is a zero mode locally bound to each quasihole. Instead the zero-mode 
wave functions are shared among the quasiholes. 

We should point out that Greiter et al.\ \cite{gww} also stated that in the
presence of quasiholes there are wave functions with broken pairs. However, the
functions they published for states with one broken pair, both with and
without quasiholes (see eqs.\ (9) and (10) in the first paper in Ref.\
\cite{gww} and (6.2), (6.8) and (6.9) in the second), vanish identically when
antisymmetrized. Probably for this reason, the counting of the number of
unpaired fermions that can be accomodated ``naturally'', i.e.\ in zero-energy
eigenstates, in the presence of quasiholes, is stated incorrectly to be $n$
broken pairs, when the correct answer is $n$ fermions.
Note also that the states for unpaired fermions without quasiholes 
that they give are not zero-energy eigenstates (and thus not obviously
eigenstates at all). 

Linear independence of the states (\ref{pf+2nqholes}) is easily established, 
for fixed $w$'s. After removing the
factor $\prod(z_i-z_j)^q$, we arbitrarily divide the particles into pairs and 
let the members of each pair approach each other, one after the other, say
$z_1\rightarrow z_2$, then $z_3\rightarrow z_4$, and so on. For each limit we
examine the leading behavior; clearly the leading behavior for each limit 
may be a single pole, in view of the paired form of the function, or it may be
nonsingular. If it is a pole, we take the function that multiplies the pole 
(its residue), which is a function of the remaining coordinates to which the 
procedure has not yet been applied, as well as of the coordinates 
$z_1=z_2$, $z_3=z_4$, \ldots, to which it has, and we repeat the process. If
the first limit is nonsingular, we call the function we started with 
the zeroth nonsingular residue; if the first limit is singular, but the 
second is not, we call the first residue the first nonsingular residue, 
and so on. Thus the $m$-th residue, obtained after the $m$-th limit, 
may be singular or nonsingular in the next limit; if it is nonsingular 
we can identify the original states as having $m=(N-F)/2$ unbroken pairs. 
Then the $(N-F)/2$-th nonsingular residue will, by definition, 
be a nonsingular function of the paired coordinates $z_1=z_2$, $z_3=z_4$, 
\ldots, $z_{N-F-1}=z_{N-F}$, the unpaired
coordinates $z_{(N-F)+1}$, \ldots, $z_N$, and of $w_1$, \ldots, $w_{2n}$.
Since two functions, one of which is singular and the other nonsingular in a
given limit are linearly independent of each other, it follows 
that states with different numbers $F$ of unpaired particles 
are linearly independent. For states with the same $F$, we consider the 
$(N-F)/2+1$-th residue, as a function of the remaining
$F$ coordinates $z_{(N-F)+1}$, \ldots, $z_N$. It is just a Slater determinant 
in these variables, and these determinants for distinct sets of $m_k$ are 
obviously linearly independent. This concludes the proof. 

{}From the wave functions (\ref{pf+2nqholes}), it is straighforward to enumerate
the number of states that satisfy the conditions, for given positions of the
$w$'s. First we note that for $0$ quasiholes, a state (the original Pfaffian 
ground state) exists for $N$ even, but not for $N$ odd. For $2n=2$ quasiholes, 
there is a unique possibility, both for $N$ even (with $F=0$) and $N$ odd 
(with $F=1$ fermion, in the $m_1=0$ state). For $2n=4$, there are $2$ states
both for $N$ even and odd; for $N$ even these are the same as the two
independent states found before. In general, for given $n>0$ and $F$, 
there are clearly
\begin{equation}
\left(\begin{array}{c}
n\\
F\end{array}\right)
\end{equation}
independent states. Summing over the allowed values of $F$, which are those
with the same parity as $N$, we obtain, whether $N$ is even or odd, $2^{n-1}$,
by a well-known formula for binomial coefficients. This number, which is valid
for $N\geq n$, is exactly the
number of conformal blocks for $2n$ spin fields in the Majorana conformal field
theory \cite{mr}; see also Ref.\ \cite{nayak}. 

Just as for the quasiholes of the Laughlin state (see Sec.\ I), there is a 
(finite) positional
degeneracy associated with the positions of the quasiholes.
The functions for fixed $w$'s are analogous to coherent states formed out of
the linearly-independent quasihole states. In the present
case, this degeneracy can be calculated, for a given $F$ and set of $m_k$'s, by 
expanding all the $\Phi$'s in powers of the $w$'s:
\begin{equation}
\Phi(z_1,z_2;w_1,\ldots,w_{2n})=\frac{(2n)!}{(n!)^2}\left[z_1^nz_2^n
      +\frac{1}{2}(z_1^{n-1}z_2^n+z_1^nz_2^{n-1})e_1(-w)+
\ldots+e_{2n}(-w)\right].
\label{Phiexp}
\end{equation}
Here $e_m(-w)$ is shorthand for the elementary symmetric polynomials in the 
$w$'s, with each $w_i$ replaced by $-w_i$:
\begin{equation}
e_m(w)=\sum_{i_1<i_2<\cdots<i_m}w_{i_1}w_{i_2}\ldots w_{i_m}.
\end{equation}
which arise since each $w$ appears at most 
once in any term resulting {}from the expansion of $\Phi$. It is known that
linear combinations of products of the elementary symmetric polynomials $e_m$, 
$m=0$, \ldots, $2n$ yield all the symmetric polynomials in $2n$ variables. 
Thus, when the functions in eq.\ (\ref{pf+2nqholes}) are expanded in powers of
the $w$'s, we obtain all the symmetric polynomials in 
$w_1$, \ldots, $w_{2n}$, in which the degree in any one $w$ is not greater than
$(N-F)/2$. The total number of linearly-independent states (as functions of the
$z$'s), for a fixed $F$ and fixed set of $m_k$'s, cannot be greater than the 
number of linearly-independent symmetric functions of the $w$'s obtained in
this expansion. Notice that, if the $w$'s are regarded as the
coordinates of some kind of particles, the symmetric polynomials in the
$w$'s can be interpreted as the states for $2n$ bosons each of which
can occupy one of $(N-F)/2+1$ orbitals (the orbitals having the single-particle
wave functions $1$, $w$, \ldots, $w^{(N-F)/2}$; they form an angular momentum
multiplet of angular momentum $(N-F)/4$). The number of such symmetric 
polynomials in the $w$'s for the Pfaffian case is thus
\begin{equation}
\left(\begin{array}{c}
(N-F)/2+2n\\
2n
\end{array}\right).
\end{equation}
It is our claim, that for each of these linearly-independent symmetric 
polynomials we have a linearly-independent many-particle 
state (for each set of $m_k$'s), and so the upper bound just obtained is in
fact the answer. To establish the truth of this claim, we again
make use of the residues of the successive limits $z_1\rightarrow z_2$,
$z_3\rightarrow z_4$, \ldots, as defined above. The nonsingular residues have a
simple form; they are proportional to
\begin{equation}
\prod_{i=1}^{(N-F)/2}\prod_{r=1}^{2n}(z_{2i}-w_r)\sum_{\sigma\in S_F}
{\rm sgn}\,\sigma \prod_{k=1}^F z_{\sigma(N-F+k)}^{m_k}.
\end{equation}
The last factor, involving the unpaired particles, is a Slater determinant; the
first factor is simply a product of Laughlin-like quasiholes acting on the
coordinate of each pair. We have thus reduced the analysis to the case of the
Laughlin states, where all the quasihole states are linearly independent (see
Sec.\ I), and this establishes our claim.

The total number of linearly-independent quasihole states, fixing only 
$N$ and $n$, is then
\begin{equation}
\sum_{F,\,(-1)^F=(-1)^N}
  \left(\begin{array}{c}
n\\
F\end{array}\right)
\left(\begin{array}{c}
(N-F)/2+2n\\
2n
\end{array}\right).
\end{equation}              
(Notice that this expression incorporates both restrictions $F\leq n$, $F\leq
N$.) This number is clearly larger than the number that would be expected if the
quasiholes behaved like the quasiholes of the Laughlin states. In the
present case the expectation, based on assuming abelian fractional statistics
as for the Laughlin states, would be (in view of the half flux in each
quasihole) that the number would be given by the formula for $2n$ bosons which
may each occupy any of $N/2+1$ orbitals (for $N$ even). We can compare our 
result with this number, which is 
\begin{equation}
\left(\begin{array}{c}
N/2 +2n\\
2n
\end{array}\right).
\end{equation}
For $n$ fixed and $N$ tending to infinity, the ratio tends to $2^{n-1}$. Again,
this represents the degeneracy necessary for nonabelian statistics. 

We now give arguments that the states found above are a complete set of zero
energy states. The general
construction of zero-energy states for the three-body Hamiltonian for $q=1$, or
its generalizations for $q>1$, was given in
the first part of Appendix A of Ref.\ \cite{milr}. It shows that without loss
of generality, zero-energy states are linear combinations of the forms (from
which we have removed the ubiquitous factor $\prod(z_i-z_j)^q$)
\begin{equation}
\sum_{\sigma\in S_N} {\rm sgn}\,\sigma
\frac
{\sum_{\tau\in S_{N/2}}\prod_{k=1}^{N/2}f_k(z_{\sigma(2\tau(k)-1)},
z_{\sigma(2\tau(k))})}
{ (z_{\sigma(1)}-z_{\sigma(2)})\cdots(z_{\sigma(N-1)}-z_{\sigma(N)}) },
\label{genzeroen}
\end{equation}
where the $f_k$ are symmetric polynomials in two variables. For $N$ odd we can
write a similar form with $k=1$, \ldots, $(N-1)/2$, and include for the unpaired
particle an arbitrary polynomial factor $f_0(z_{\sigma(N)})$. In order to 
represent states with $2n$ quasiholes, the number of flux added to the
Pfaffian ground state must be $n$, and so the $f_k$ (and $f_0$ for $N$ odd) 
must be of degree at most $n$ in each coordinate $z_i$; these symmetric
polynomials must then be linear combinations of the linearly-independent forms
\begin{equation}
z_1^{n_1}z_2^{n_2}+z_1^{n_2}z_2^{n_1},
\label{symmpols}
\end{equation}
in which $0\leq n_1\leq n_2\leq n$. These clearly span a vector space of 
dimension $\frac{1}{2}(n+1)(n+2)$.
As in Ref.\ \cite{milr}, if any $f_k(z_1,z_2)$ vanishes at $z_1=z_2$, then it
must contain a factor $(z_1-z_2)^2$, and this pair is broken and will
contribute to the unpaired-fermion part of the functions
(\ref{pf+2nqholes}). The subspace of symmetric polynomials for which this
occurs is spanned by $(z_1-z_2)^2$ times those in eq.\ (\ref{symmpols}), 
here with $0\leq n_1\leq n_2\leq n-2$; this subspace has dimension 
$\frac{1}{2}n(n-1)$. The quotient of these two spaces, which
represents the symmetric polynomials in two variables which do not vanish at 
$z_1=z_2$, therefore has dimension $2n+1$. But we have already found a set of 
such functions while expanding $\Phi$, eq.\ (\ref{Phiexp}), and there are 
exactly $2n+1$ terms in this series, which form a linearly-independent set for 
the required symmetric polynomials. Therefore we may now argue that, in the 
general form (\ref{genzeroen}), we may choose the $f_k$ to be in either of the 
two sets, that is those that vanish at $z_1=z_2$ and those appearing in eq.\
(\ref{Phiexp}). The unpaired fermion part takes the given form, after use 
of the antisymmetrization among the $F$ particles involved. If $f_k$ does not 
vanish at $z_1=z_2$, then it is part of the expansion of $\Phi$ for that pair. 
Thus we get exactly all the terms that occur on expanding 
eq.\ (\ref{pf+2nqholes}).
This shows that the count of states given is correct, and that they can 
be viewed as arising {}from the states with quasiholes at fixed positions.

The numbers for the total number of zero-energy states, which are not resolved 
into angular momentum multiplets, are convenient for comparison with numbers 
of zero-energy eigenstates obtained numerically (using the same three-body 
Hamiltonian (\ref{pfaff3bodprojH})), when these are summed over all $L$ and 
$L_z$. We can also work out 
the decomposition of the functions (\ref{pf+2nqholes}) into angular momentum 
multiplets, using the wave functions (\ref{pf+2nqholes}) and applying the 
Clebsch-Gordan decomposition to the angular momenta of 
the bosons that represent the quasiholes in the $\Phi$ factors, and the 
unpaired fermions. These numbers, and the angular momentum decomposition, are 
in perfect agreement for moderate sizes. The cases that have been checked 
numerically are shown in Table \ref{tab:pfzeroen}. In the Table, we have shown
only the results for $N$ even; similar results, in complete agreement with the
analytic formulas, are also found for $N$ odd.        

As an aside, it is interesting that in this case we have established the linear
independence of the functions in general, whereas in Ref.\ \cite{milr} we were 
forced to resort to a case-by-case analysis. In fact, the results given here 
now suffice to complete the proof of linear independence of the edge states of 
the Pfaffian state on a disk. We first note that if we place many (or all) 
quasiholes at the same place by setting all the $w$'s equal, then there are no 
particles in that region of the sphere, and the fluid has an edge there. If we 
take the limit of these states as $N_\phi\rightarrow\infty$ with $N$ fixed, 
then the sphere becomes the infinite plane, and the particles are concentrated 
in a disk at the origin if the quasiholes are at the position on the sphere 
that is mapped to the point at infinity by the stereographic projection. Thus, 
the problems of finding the zero-energy quasihole and edge states are 
essentially the same. The general edge states, that include charge-fluctuation 
excitations at the edge, are obtained by letting some of the $w$'s deviate 
{}from infinity, and expanding in $1/w$'s gives states that contain symmetric 
polynomials in the paired-particle coordinates, rather than in all the 
coordinates as in Ref.\ \cite{milr}. This is merely another basis for the edge 
states; the number of states at each angular momentum level is easily seen to 
be the same, for sufficiently large $N$. In the limit, the complete and 
linearly-independent zero-energy bulk quasihole states, for the case where all 
$w$'s are nearly equal, yield all the edge states of the disk, and this gives 
a proof that the latter are linearly independent. To obtain results for the 
cylinder, with two edges, we place half the quasiholes at each pole of the 
sphere and take a similar limit. In the limit, the particles occupy a narrow 
band around the equator of the sphere, and the infinite flux through the 
system makes it equivalent to the cylinder, if we consider states where the 
particles are close to the equator, which again means the $w$'s must not 
deviate far {}from the poles. Note that the fact that the particles are spread 
very thinly along this band in the limit we have taken does not affect the 
construction or counting of edge states, which is independent of the aspect 
ratio of the occupied region of the cylinder. Also, the operation that shifts 
charge {}from one edge to the other \cite{milr} is obtained by removing a 
quasihole {}from one pole and placing it at the other.

In Fig.\ \ref{fig:pf0qh} we show numerical spectra for the 
$q=2$ three-body Hamiltonian, described after (\ref{pfaff3bodprojH}) (in which
the projection is onto $L=3N_\phi/2-3$), 
with $V=1/6$, for $N=12$ electrons and $N_\phi=2N-3$ flux, that is, 
no quasiholes, and $\nu=1/2$. In Fig.\ \ref{fig:pf4qh} we show the same 
but with $N=10$ and $2$ flux added, so there are $2n=4$ quasiholes. The 
zero-energy states, of which the degeneracies were given in Table
\ref{tab:pfzeroen}, can be 
seen at $E=0$, as can the set of angular momentum values obtained in this case.
The figure also shows that all higher energy states are separated by a 
significant gap that we expect will survive in the thermodynamic limit, as 
needed for the arguments for nonabelian statistics. To further substantiate the
claim that this system is incompressible, we include some finite size scaling
results. In Fig.\ \ref{fig:pfneutsize}, we show the low-lying excited energy 
levels above the ground state for $N=10$ and $12$, versus $k=L/R$. It can be 
seen that the spectra lie almost on top of one another. In 
Fig.\ \ref{fig:pfneutgapscal}, we show the size dependence of the lowest excited
state, versus $1/N$, for several values of $N$, including those in Fig.\
\ref{fig:pfneutsize}. Although it is not quite clear by inspection that this 
energy gap is converging to a constant as $N\rightarrow \infty$, because of the
behavior of the points for the sizes $N=10$, $12$, we note that in
Fig.\ \ref{fig:pfneutsize} the lowest-energy points for $N=10$ seem to lie on 
two sides of a minimum, while for $N=12$ one lies near a minimum. Hence we 
believe that these results do converge to a finite gap. In Fig.\ 
\ref{fig:pfqelgapscal}, we show the ground state energy for systems with 
$N_\phi=2(N-1)-2$, that is two {\em quasielectrons}, where there are no 
zero-energy states for the three-body Hamiltonian. These energies do appear 
to be converging to a finite gap as $N\rightarrow\infty$. This energy 
determines the slope in the total energy density versus density in the 
thermodynamic limit, on the higher-density side of $\nu=1/2$. Since the energy 
density is zero on the lower-density side, this represents a discontinuity in 
the chemical potential at $\nu=1/2$ for this Hamiltonian, showing that the 
system is incompressible.

\section{Quasiholes of the Haldane-Rezayi state on the sphere}
\label{HR}

The Haldane-Rezayi (HR) state \cite{haldrez} can be written in terms of the 
coordinates of $N/2$ up-spin particles at $z_1^\uparrow$, \ldots, and $N/2$ 
down-spin particles at $z_1^\downarrow$, \ldots, as   
\begin{eqnarray}      
\lefteqn{\tilde{\Psi}_{\rm HR}(z_{1}^{\uparrow},\ldots,z_{N/2}^{\uparrow},
z_1^{\downarrow},\ldots,z_{N/2}^{\downarrow})=}\nonumber \\
   & & \sum_{\sigma\in S_{N/2}}{\rm sgn}\,\sigma
         \frac{1}{(z_{1}^{\uparrow}-z_{\sigma(1)}^{\downarrow})^{2}
           \cdots(z_{N/2}^{\uparrow}-z_{\sigma(N/2)}^{\downarrow})^{2}}
\prod_{i<j}(z_{i}-z_{j})^{q}.
\label{HRground}
\end{eqnarray}           
Here $q\geq2$ is even for electrons, and odd for bosons, and the filling 
factor is $1/q$. The first factor is of course just a determinant. The product 
over $z_i$'s with no spin labels attached is over all particles.
The fact that this describes a singlet is discussed carefully in \cite{haldrez}.
In \cite{mr} it was pointed out that this state can be regarded as a BCS-type 
condensate of spin-singlet pairs of spin-1/2 composite fermions that consist of 
a particle and $q$ vortices, {}from which the spin-singlet property can be more
easily understood. 
Some further discussion related to the edge states is contained in \cite{milr}.
The HR state is the unique zero-energy state at $N_\phi=q(N-1)-2$ flux of a
``hollow-core'' pseudopotential Hamiltonian that gives any two particles a
nonzero-energy when their relative angular momentum is either $q-1$ or $\leq
q-3$ \cite{haldrez}, again with the assumption that the particles are bosons 
for $q$ odd, fermions for $q$ even, and that $q\geq 2$.

In exact analogy with the Pfaffian state, the 
wave function for two quasiholes is:
\begin{eqnarray}
\lefteqn{\tilde{\Psi}_{\rm HR}(z_{1}^{\uparrow},\ldots,z_{N/2}^{\downarrow};
w_{1}, w_{2})=}\nonumber \\
  & & \sum_{\sigma\in S_{N/2}}{\rm sgn}\,\sigma
            \frac{\prod_{k=1}^{N/2} 
             [(z_{k}^{\uparrow}-w_{1})(z_{\sigma(k)}^{\downarrow}-w_{2})
                          +(w_{1}\leftrightarrow w_{2})] }
   {(z_{1}^{\uparrow}-z_{\sigma(1)}^{\downarrow})^{2}\cdots
    (z_{N/2}^{\uparrow}-z_{\sigma(N/2)}^{\downarrow})^{2}} 
\prod_{i<j}(z_{i}-z_{j})^{q}.
\end{eqnarray}           
Due to the spin-independence of the newly-inserted factors acting on each pair
inside the sum over permutations, the state is still a spin-singlet and this 
suggests that the quasiholes carry no spin. The two-quasihole state is again 
a zero-energy eigenstate of the hollow-core Hamiltonian. To see this fact, 
expand the inserted factors for each pair in terms of powers of 
$z_k^\uparrow \pm z_{\sigma(k)}^\downarrow$. Due to the symmetry between 
$z_k^\uparrow$ and $z_{\sigma(k)}^\downarrow$ in each factor, it is easy to 
see that $z_k^\uparrow-z_{\sigma(k)}^\downarrow$ must occur to an even power. 
Thus in the complete wave function, the absence of 
$(z_k^\uparrow-z_{\sigma(l)}^\downarrow)^{q-1}$ for any $k$, $l$, and hence the
zero-energy property of the ground state, is preserved in the quasihole states. 

It is possible to write down directly the forms of all the zero-energy states
of the hollow-core Hamiltonian, in analogy with those for the Pfaffian and
those in Ref.\ \cite{milr}. In terms of the coordinates of $N_\uparrow$ 
up particles, $N_\downarrow$ down particles, the wave functions are linear 
combinations of
\begin{eqnarray}
&&\frac{1}{(N_\uparrow-F_\uparrow)!}
\sum_{\stackrel{\scriptstyle\sigma\in S_{N_\uparrow}}{\rho\in
S_{N_\downarrow}}}
{\rm sgn}\,\sigma \,{\rm sgn}\,\rho    
\prod_{k=1}^{F_\uparrow} (z_{\sigma(k)}^\uparrow)^{n_k}
 \prod_{l=1}^{F_\downarrow} (z_{\rho(l)}^\downarrow)^{m_l}
\prod_{r=1}^{N_\uparrow-F_\uparrow}
\frac{\Phi(z_{\sigma(F_\uparrow+r)}^\uparrow,
z_{\rho(F_\downarrow+r)}^\downarrow;w_1,\ldots,w_{2n})}
{(z_{\sigma(F_\uparrow+r)}^\uparrow -
z_{\rho(F_\downarrow+r)}^\downarrow)^2}\nonumber\\
&&\qquad\qquad\qquad\times\prod_{i<j}(z_i-z_j)^q.
\label{hr+2nqholes}
\end{eqnarray}              
Here $N_\uparrow-F_\uparrow=N_\downarrow-F_\downarrow$ is the number of
unbroken pairs, and we may assume the $n_k$'s, $m_k$'s are strictly increasing,
as for those in the Pfaffian quasihole states, with $0\leq
n_1<n_2<\cdots<n_{F_\uparrow}\leq n-2$, 
$0\leq m_1<m_2<\cdots<m_{F_\downarrow}\leq n-2$; consequently, $0\leq F_\sigma
\leq n-1$ for each $\sigma=\uparrow$ or $\downarrow$. 
As written, these states do not 
have definite spin, but
eigenstates of ${\bf S}^2$ and of $S_z$ can be constructed.
Since the paired particles form singlets, the spin is determined by the spin
1/2 unpaired fermions in the sums over $\sigma$ and $\rho$, which behave
identically to ordinary spin 1/2 fermions. Hence the possible spin states are
determined by adding the spins of particles in different orbitals (labelled by
$n_k$ or $m_k$), with the only constraint that an orbital occupied with both an
up and a down fermion must form a singlet. The total spin $S$ of the zero-energy
states therefore obeys $S\leq (n-1)/2$. Notice that the number of flux in
these states is $N_\phi=q(N-1)-2+n$.
Arguments that the states given are both complete and linearly 
independent, so that our count of states is correct, can be constructed 
straightforwardly as a combination of those in Appendix A of Ref.\ \cite{milr} 
and in Sec.\ \ref{pfaffian} above; we omit the details.

We may now count the number of linearly-independent states as for the Pfaffian.
We see that for $n=0$, $1$, we must have $F_\uparrow=F_\downarrow=0$, and such
states exist only for $N=N_\uparrow+N_\downarrow$ even; they are the 
ground and two-quasihole states written down above. For $n=2$, we find two
possibilities for both $N$ even or odd, like the Pfaffian case. For the general
case, we can write the number of states for fixed $n$,
$F_\uparrow$, $F_\downarrow$, and $w$'s in two ways. One is 
\begin{equation}
\left(\begin{array}{c}
n-1\\
F_\uparrow\end{array}\right)
\left(\begin{array}{c}
n-1\\
F_\downarrow\end{array}\right).
\end{equation}
If we sum over $F_\uparrow$ with $F=F_\uparrow+F_\downarrow$ fixed, we obtain a
second form,
\begin{equation}
\left(\begin{array}{c}
2(n-1)\\
F \end{array}\right),
\end{equation}
which is the number of states for $F$ fermions in $2(n-1)$ orbitals. The sum
over $F$ satisfying $(-1)^F=(-1)^N$ then yields a total of
$2^{2(n-1)-1}=2^{2n-3}$ zero-energy quasihole states of all spins for fixed $N$
(either even or odd, except for small $n$ as already shown) and $w$'s. 
                                                             
We now find the dimension of the space of quasihole states at all spins for
fixed $N$, including the positional degeneracy due to the $w$'s as for the
Pfaffian. Then the total number of linearly-independent 
quasihole states, fixing only $N$ and $n$, is
\begin{equation}
\sum_{F,\,(-1)^F=(-1)^N}
  \left(\begin{array}{c}
2(n-1)\\
F\end{array}\right)
\left(\begin{array}{c}
(N-F)/2+2n\\
2n
\end{array}\right).
\end{equation}              
Again the ratio of this number to that for positional degeneracy only is
$2^{2n-3}$ as $N\rightarrow\infty$. In this case, the factor $2^{2n}$ might
give the impression that there is a factor 2 for each quasihole, perhaps
because each carries spin $1/2$. But the result is in fact $2^{2n-3}$, which
indicates the connection with nonabelian statistics. There are only $n-1$ zero
modes available, similarly to the Pfaffian, which can be occupied with either 
spin, with a final condition on the parity of the number of unpaired fermions.

The two-particle correlation functions for the HR ground state with $q=2$ 
have been published previously \cite{haldrez}, for $6$ particles; they suggest 
that the correlation length is quite large in this sytem also.               
In Table \ref{tab:hrzeroen} we show results obtained numerically for 
zero-energy quasihole
states of the hollow-core Hamiltonian for $q=2$, which agree exactly with the 
general formula, as do the orbital and spin angular momenta. 
In Fig.\ \ref{fig:hr0qh}, we show the spectrum of the HR state at $\nu=1/2$, 
when no quasiholes are present, for $N=8$ particles. 
In the spectra, we have chosen $V_1=1$.
There does appear to be a
gap above the ground state, and the form of the low-lying excitation spectrum
is similar to that of the Pfaffian.
In Fig.\ \ref{fig:hr8qh}, we show the spectrum when $8$ quasiholes are 
present, for $N=6$ and $N_\phi=12$. However, we note that because the 
particles carry spin in this system, it is harder to get to sufficiently 
large sizes to ensure that the results are converging to the thermodynamic 
limit, and larger sizes may be needed in order to prove that the energy gaps 
are approaching constants in this limit.

\section{Quasiholes of the 331 state on the sphere}
\label{331}

It is of interest to perform a similar calculation for the quasiholes of the 331
state, even though the excitations of this state are known to have abelian
statistics. The relation to paired states has been discussed in
\cite{gww2,halpnewport,ho,milr}. 

The 331 state is just one of a family of two-component states, the so-called
$mm'n$ states, first introduced by Halperin \cite{halp83}, and studied further
in Refs.\ \cite{hald,hr87,ymg2,he,gww2}. 
Using notation $\uparrow$, $\downarrow$ for the two components, 
these states can be written:
\begin{equation}
\tilde{\Psi}_{mm'n}(z_1^\uparrow,\ldots,z_{N_\uparrow}^\uparrow,
z_1^\downarrow,\ldots,z_{N/2}^\downarrow)=
\prod_{i<j} (z_i^\uparrow-z_j^\uparrow)^m 
\prod_{k<l} (z_k^\downarrow-z_l^\downarrow)^{m'} 
\prod_{rs} (z_r^\uparrow-z_s^\downarrow)^n .
\label{mmnground}
\end{equation}
The general $mm'n$ state (for some values of $N_\uparrow$, $N_\downarrow$) 
is the unique lowest total-angular-momentum ground 
state of a spin-dependent pseudopotential Hamiltonian, that generalizes
(\ref{pseudH}) 
to the two-component case \cite{hald}, 
\begin{equation}
H=\sum_{i<j}\left[\sum_{M=0}^{m-1} V_M P_{ij}(N_\phi-M,\uparrow\uparrow)
+\sum_{M=0}^{m'-1} V'_M P_{ij}(N_\phi-M,\downarrow\downarrow)\right]
+\sum_{ij}\sum_{M=0}^{n-1} V''_M P_{ij}(N_\phi-M,\uparrow\downarrow)
\label{331pseudH}
\end{equation}
in which the projection operators $P_{ij}(L,\sigma\sigma')$ project onto
the spin states $\sigma$, $\sigma'$ for particles $i$, $j$, respectively, 
as well as onto total orbital angular momentum $L$. 
Thus this Hamiltonian gives positive energy to any state in which 
two $\uparrow$ or $\downarrow$ particles have relative angular momentum less 
than $m$ or $m'$, respectively, or in which an $\uparrow$ and a $\downarrow$
particle have relative angular momentum less than $n$.

For the case when the exponents in these states are of the form
$m=m'=q+1$, $n=q-1$, $q\geq1$, (which give filling factor $\nu=1/q$, and the
partial filling factors for $\uparrow$, $\downarrow$ are both $1/2q$; for 
brevity, we will continue to refer to this class of states with general
$q$ as the 331 state), 
then use of the Cauchy determinant identity,
\begin{equation}
\prod_{i<j} (z_i^\uparrow-z_j^\uparrow)
\prod_{k<l} (z_k^\downarrow-z_l^\downarrow)
\prod_{rs} (z_r^\uparrow-z_s^\downarrow)^{-1} 
 = \det \left( \frac{1}{z_i^\uparrow-z_j^\downarrow}\right),
\label{cauchident}
\end{equation}
allows the ground states to be written in a paired form (i.e., as a
spin-indepenedent Laughlin-Jastrow factor times a pairing function), 
similar to the Pfaffian and HR states \cite{haldrez,gww2}. In terms of BCS-type
pairing, this function describes $p$-type spin-triplet pairing, with each 
pair in the $S_z=0$ state of a spin triplet \cite{halpnewport,ho}.

We will write the quasihole states immediately in terms of broken pairs, 
(although a simple form using Laughlin quasihole operators acting on either
$\uparrow$ or $\downarrow$ spins will be described later):
\begin{eqnarray}
&&\frac{1}{(N_\uparrow-F_\uparrow)!}
\sum_{\stackrel{\scriptstyle\sigma\in S_{N_\uparrow}}{\rho\in
S_{N_\downarrow}}}
{\rm sgn}\,\sigma \,{\rm sgn}\,\rho
\prod_{k=1}^{F_\uparrow} (z_{\sigma(k)}^\uparrow)^{n_k}
\prod_{l=1}^{F_\downarrow} (z_{\rho(l)}^\downarrow)^{m_l}
\prod_{r=1}^{N_\uparrow-F_\uparrow}
\frac{\Phi(z_{\sigma(F_\uparrow+r)}^\uparrow,
z_{\rho(F_\downarrow+r)}^\downarrow;w_1,\ldots,w_{2n})}
{z_{\sigma(F_\uparrow+r)}^\uparrow -
z_{\rho(F_\downarrow+r)}^\downarrow}\nonumber\\
&&\qquad\qquad\qquad\times\prod_{i<j}(z_i-z_j)^q
\label{331+2nqholes}
\end{eqnarray}              
which is particularly similar to the HR case, except that here $n_k$ and
$m_l\leq n-1$, and so $0\leq F_\sigma\leq n$. The flux in these states is 
$N_\phi=q(N-1)-1+n$, as for the Pfaffian.

For the count of states at fixed $w$'s we obtain, again summing over 
$F_\uparrow$ with $F=F_\uparrow+F_\downarrow$ fixed, 
\begin{equation}
\left(\begin{array}{c}
2n\\
F \end{array}\right),
\end{equation}
which is the number of states for $F$ fermions in $2n$ orbitals. The sum
over $F$ satisfying $(-1)^F=(-1)^N$ then yields a total of
$2^{2n-1}$ zero-energy quasihole states of all spins for fixed $N$
and $w$'s. This is valid for all $n$ except $n=0$, in which case there is no
zero-energy state for $N$ odd. 

The result $2^{2n-1}$ may also be understood by viewing it as
coming {}from the choice of layer index on the Laughlin quasihole operator.
Thus quasihole states of zero energy can clearly be obtained by multiplying
factors $\prod(z_i^\sigma-w^\sigma)$, that act on either particles of spin
$\sigma=\uparrow$ only, or on spin $\downarrow$ only, into the 331 ground
state. Further, the numbers $N_\uparrow$, $N_\downarrow$ need not be equal.
However, the flux seen by both $\uparrow$ and $\downarrow$ particles must be
equal. This leads to the condition
$2(N_\uparrow-N_\downarrow)=n_\downarrow-n_\uparrow$, where $n_\sigma$ are the
numbers of quasiholes of the two types, in which the factors act on the 
particles of spin $\sigma$. For fixed $N=N_\uparrow+N_\downarrow$, there are
$2^{2n-1}$ ways to choose the spins of the quasiholes consistent with these
conditions, for $n\geq 1$. We will refer to this construction of the
zero-energy states as bosonic, because, as explained in Ref.\ \cite{milr}, the
relation of the two approaches is related to bosonization of fermi systems.

Returning to the paired, or fermionic, description, the total number of states 
resulting {}from the positional degeneracy is in this case
\begin{equation}
\sum_{F,\,(-1)^F=(-1)^N}
  \left(\begin{array}{c}
2n\\
F\end{array}\right)
\left(\begin{array}{c}
(N-F)/2+2n\\
2n
\end{array}\right).
\end{equation}              
This is larger than for the Pfaffian or HR states. It should also be possible
to obtain this formula {}from the bosonic description, although we have not done
so. The positional degeneracy for the quasiholes of each spin is in that
approach
\begin{equation}
\left(\begin{array}{c}
N_\sigma+n_\sigma\\
n_\sigma\end{array}\right),
\end{equation}
and the summations are constrained by the fact that for the zero-energy states 
$N_\sigma$ depends on the $n_\sigma$.

For the 331 state, it is again not too difficult to extend the arguments for
the Pfaffian or HR cases, to show the completeness and linear independence of
the zero-energy states found, in the pairing or fermionic form. This is also 
quite clear in the bosonic form of the states.                      

\section{Ground states on the torus}
\label{torus}

In this section we consider the zero-energy eigenstates of the special
Hamiltonians discussed above on a system with periodic boundary conditions (a
torus), without quasiholes. Although we believe that the states we will give
span the complete spaces of such states, we will not prove this, but will refer
to numerical results for confirmation.

First we briefly review known results for this geometry \cite{hr85}, 
to fix notation. In
the Landau gauge ${\bf A}=-By\hat{\bf x}$, we will take the magnetic length to
be 1, and the system to be a parallelogram with sides $L_1$, $L_2$, and 
periods $L_1$, $L_2e^{i\alpha}=L_1\tau$ in the complex plane; $\alpha$ is the
angle between the sides, and $\tau$ (with ${\rm Im}\, \tau>0$) parametrizes 
the aspect ratio. As usual, there are $N_\phi$ flux through the surface.
Many-particle wave functions in the LLL can be 
written 
\begin{equation}
\Psi(z_1,\ldots,z_N)=f(z_1,\ldots,z_N) e^{-\sum_i y_i^2/2},
\end{equation}
where $f$ is a holomorphic function; as a consequence of the boundary
conditions on $\Psi$ it is required to satisfy, for all $i$,
\begin{eqnarray}
\frac{f(z_1,z_2,\ldots,z_i+L_1,\ldots)}{f(z_1,z_2,\ldots,z_i,\ldots)}&=&
e^{i\phi_1},     \nonumber\\
\frac{f(z_1,z_2,\ldots,z_i+L_1\tau,\ldots)}{f(z_1,z_2,\ldots,z_i,\ldots)}&=&
e^{i\phi_2} e^{-i\pi N_\phi(2 z_i/L_1+\tau)}.
\label{perbcs}
\end{eqnarray}
Here $\phi_1$, $\phi_2$ represent general twisted boundary conditions, the same
for all particles, and are the same for all states in the Hilbert space. They
can be set to zero without any real loss of generality.

{}From a general symmetry analysis \cite{hald85}, any state in the system can be
decomposed into center of mass and relative motion, as 
\begin{equation}
f(z_1,\ldots,z_N)=F_{\rm cm}(Z) f_{\rm rel}(z_1,\ldots,z_N),
\label{cmdecomp}
\end{equation}
where $f_{\rm rel}$ is invariant under shifts of all $z_i$ by the same amount,
and $F_{\rm cm}$ is a function of $Z=\sum_i z_i$ only. Given the boundary
conditions on $f$, specified by $\phi_1$, $\phi_2$, there is still some 
freedom of choice in the corresponding phases in the conditions on $F_{\rm cm}$
and $f_{\rm rel}$, which is related to Haldane's $\bf k$-vector quantum number 
\cite{hald85},
and will be useful in the following. In any case, there are always $q$
solutions for $F_{\rm cm}$, for filling factor $\nu=p/q$ ($p$, $q$ coprime),
which are mapped into each other by magnetic translations of the center of 
mass, and consequently are degenerate in energy for any 
translationally-invariant Hamiltonian.

For the Laughlin state on the torus, the property of vanishing as the $q$th
power as any $z_i\rightarrow z_j$ fixes the relative wave function to be 
(definitions and results for the theta and elliptic functions used in this 
section are given in Appendix A)
\begin{equation}
f_{\rm rel}=\prod_{i<j}\vartheta_1((z_i-z_j)/L_1|\tau)^q\equiv f_{\rm LJ}.
\end{equation}
The basis states for the center-of-mass wave functions found in Ref.\ 
\cite{hr85} can be rewritten
(see Appendix A),
apart {}from some constant factors, as \cite{krunpub}
\begin{equation}
F_{\rm cm}(Z)=\vartheta\left[\begin{array}{c} 
                   \ell/q+(N_\phi-q)/2q+\phi_1/2\pi q\\
                      -(N_\phi-q)/2-\phi_2/2\pi  \end{array}\right]
                          (qZ/L_1|q\tau).
\label{Fcm}
\end{equation}
Here $\ell=0$, $1$, \ldots, $q-1$ labels the center-of-mass degeneracy, and
$\phi_1$, $\phi_2$ have been retained for generality. 
Using the properties given in App.\ \ref{theta}, this can be verified to obey
the conditions resulting {}from (\ref{perbcs}) that are given in \cite{hr85}.
Alternatively one can verify directly that eq.\ (\ref{cmdecomp}) satisfies eq.\
(\ref{perbcs}). Note that $F_{\rm cm}$
has $q$ zeroes in the unit cell for $Z$ of sides $L_1$, $L_1\tau$, and linear
combinations of these functions span the Hilbert space of a charged particle on
a torus in a magnetic field with $q$ flux quanta through the torus, in the LLL.
The flux seen by the particles in these states is $N_\phi=qN$. 

For the paired states on the torus, the relative motion part $f_{\rm rel}$ must
be modified {}from $f_{\rm LJ}$, in particular to change the symmetry under
permutations, which can be done at the same flux $N_\phi=qN$, by using
\begin{equation}
f_{\rm rel}=f_{\rm elliptic} f_{\rm LJ}.
\end{equation}
Here $f_{\rm elliptic}(z_1,\ldots,z_N)$ is a meromorphic function 
and, since $f_{\rm rel}$ must be holomorphic, any poles in $f_{\rm elliptic}$
must be of sufficiently low order that they are rendered nonsingular by the 
zeroes of $f_{\rm LJ}$, which are located on the hyperplanes $z_i=z_j$.
Further, $f_{\rm elliptic}$ must obey
$z_i$-independent boundary conditions
\begin{equation}
f_{\rm elliptic}(z_1,\ldots,z_i+L_1,\ldots,z_N)\propto 
f_{\rm elliptic}(z_1,\ldots,z_i,\ldots,z_N)
\label{fellipbc}
\end{equation}
and similarly for $z_i\rightarrow z_i+L_1\tau$. 
Thus $f_{\rm elliptic}$ is analogous to an elliptic function, but in 
$N$ complex variables.
In general, the phases that are
the proportionality factors in (\ref{fellipbc}) need not be equal to 1, because
any phase left by translating $f_{\rm rel}$ can be absorbed (cancelled) by
modifying the behavior of $F_{\rm cm}$, by a shift in $\phi_1$ or $\phi_2$ (by
$\pi$ in the present case) in 
eq.\ (\ref{Fcm}) before setting $\phi_1$ and $\phi_2$ to zero. This is the
freedom of choice that is related to the $\bf k$-vector, mentioned above.
For the paired states, $f_{\rm elliptic}$ is expected to be a periodic
generalization of the pairing functions, such as the Pfaffian, discussed
earlier on the sphere.

We now review the ground states of the three-body Hamiltonian eq.\ 
(\ref{pfaff3bodH}) and its generalizations on the torus. These
ground states, without quasiholes, were found by Greiter et al.\ \cite{gww}. 
For the relative motion part we take
\begin{equation}
f_{{\rm elliptic},a}={\rm Pf}\,\left[\frac{\vartheta_a((z_i-z_j)/L_1|\tau)}
{\vartheta_1((z_i-z_j)/L_1|\tau)}\right]
\end{equation}
where $a=2$, $3$, or $4$. (The three ratios $\vartheta_a/\vartheta_1(z|\tau)$ 
are essentially the three Jacobi elliptic functions ${\rm sn}\,(z|\tau)$, 
${\rm cn}\,(z|\tau)$, and ${\rm dn}\,(z|\tau)$, up to translations of $z$ and
some factors.) These then reverse sign under exactly two of the
three transformations $z_i\rightarrow z_i+L_1$, $z_i\rightarrow z_i+L_1\tau$, 
$z_i\rightarrow z_i+L_1(1+\tau)$, and are invariant under the third, for any 
$i$. The change in sign can be absorbed by the effect of a modification of
$F_{\rm cm}$ as explained above, such that the full wave function always obeys
the same boundary conditions. This structure has the consequence that Haldane's
$\bf k$-vector quantum number \cite{hald85}, which lies in a Brillouin zone, 
and which is zero (modulo reciprocal lattice vectors) in all $q$ of the 
periodic Laughlin states, is nonzero in these periodic Pfaffian states. The 
distinct nonzero values for $a=2$, $3$, or $4$, are determined 
by the behavior under the three translations already mentioned. For the most
symmetrical choice of $\tau$, $\tau=e^{i\pi/3}$, which gives the system 
six-fold rotational symmetry in real space and in the hexagonal Brillouin zone 
of $\bf k$'s, these nonzero vectors lie halfway along the shortest nonzero 
reciprocal lattice vectors; only
three of these vectors are distinct modulo addition of reciprocal lattice
vectors. Thus they lie at the midpoints of the edges of the zone, for this 
choice of zone. Thus, all $3q$ zero-energy states found for $\nu=1/q$
have distinct quantum numbers and so are linearly-independent (indeed,
orthogonal) states. Numerical calculations are in agreement with 
these quantum numbers \cite{haldpriv}.

Next we turn to the HR state on the torus. For the HR state at $\nu=1/2$ 
($q=2$), numerical calculations have revealed that there are $10=5q$ ground
states of the hollow-core model \cite{hrunpub,keski}, and so we expect to find
$5q$ for general $q$. The full set of wave functions has not to our knowledge
been obtained previously. 

Following the reasoning for the Pfaffian, we might expect $f_{\rm elliptic}$ to
be a determinant of $\vartheta_a\vartheta_b/\vartheta_1^2$'s in
$z_i^\uparrow-z_j^\downarrow$, with $a$, $b=2$, $3$, $4$, or possibly an
antisymmetrized combination of other functions that each obey the same boundary
conditions. (This assumption may be too restrictive, since the boundary
conditions in fact need only apply to the complete function. We will see that
it works for HR, but not for 331.) These products of Jacobi elliptic functions
are not, however, linearly independent. In general, elliptic functions are
completely determined by the singular part of their behavior near the poles,
and by the periodicities (see e.g.\ Ref.\ \cite{copson}). In the present case, 
we require that under
$z\rightarrow z+1$, $z\rightarrow z+\tau$ the elliptic function [with arguments
$(z|\tau)$] be either periodic or antiperiodic, giving four possibilities which
we will label $++$, $+-$, $-+$, $--$, in an obvious notation, and we also 
require that there be a double pole at $z=0$, with residue zero. 

For the $++$ case, there is a classic solution to these requirements: the
Weierstrass function $\wp(z|\tau)$. The required
functions with the other boundary conditions, which we will denote $\wp_2$, 
$\wp_3$, $\wp_4$, (in the same sequence as before), can be obtained
straightforwardly (see App.\ \ref{theta}). Defining $\wp_1=\wp$, four 
candidates for the relative 
part of the HR state on the torus are obtained:
\begin{equation}
f_{{\rm elliptic},a}=\det \wp_a((z_i^\uparrow-z_j^\downarrow)/L_1|\tau),
\end{equation}
and so, on including $f_{\rm LJ}$ and $F_{\rm cm}$, we obtain $4q$ states. It is
easy to see that they are zero-energy states of the hollow-core Hamiltonian.
The $\bf k$ values for the cases $a=2$, $3$, $4$ are as for the Pfaffian,
while $a=1$ gives states at ${\bf k}=0$.
Since the $4q$ states have distinct quantum numbers, they are linearly 
independent. 
 
To obtain the fifth set of $q$ states, we note that, for the $++$ case only, the
Weierstrass function is not the unique solution to the problem posed: we get
another solution by adding a constant. (For the other cases, this would violate
the boundary conditions.) Indeed, we could have used a constant in place of
$\wp_1$, but the determinant would then vanish except in the case of $N=2$
particles. If we insert $\wp_1+c$ in place of $\wp_1$ in the determinant, and
then expand in powers of $c$, we find that terms of order higher than one in
$c$ vanish identically because there are rows or columns in the determinant
that are equal. The term of first order, however, is linearly independent of
the zeroth order term (which is $f_{{\rm elliptic},1}$) and is nonzero. Further,
it is a zero-energy eigenstate of the hollow-core model; it clearly has ${\bf
k}=0$. Explicitly, this function is
\begin{equation}
f_{{\rm elliptic},5}=\frac{1}{(N/2-1)!}
\sum_{\stackrel{\scriptstyle\sigma\in S_{N/2}}{\rho\in
S_{N/2}}}{\rm sgn}\,\sigma \,{\rm sgn}\,\rho    
\prod_{r=1}^{N/2-1}\wp_1((z_{\sigma(1+r)}^\uparrow
-z_{\rho(1+r)}^\downarrow)/L_1|\tau).
\end{equation}     
We point out that this has an interpretation in terms of unpaired fermions.
Unpaired fermions must occupy single-particle states that are holomorphic and
obey the boundary conditions on $f_{\rm elliptic}$. For any except $++$, there
are no such states, and for $++$ the only state is, again, the constant
function. By breaking one pair in $f_{{\rm elliptic},1}$ and putting the two
fermions (with opposite spin) in this constant state, we obtain 
$f_{{\rm elliptic},5}$. We observe that all $5q$ states found are spin 
singlets. The first and fifth sets of $q$ states are linearly independent
because they have different numbers of poles.
The $\bf k$ values found agree with the numerical results.

The existence of five sets of $q$ states for the HR state on the torus is
surprising, especially as the analysis in Ref.\ \cite{milr} found just $4q$ 
sectors of edge states, and one expected \cite{mr}, on general CFT grounds,
that the number of sectors (or of primary fields of the chiral algebra of the
CFT, which comes to the same thing) would be the same as the number of
conformal blocks in the CFT on the torus. All we can say about this here is
that the CFT described in Ref.\ \cite{milr} for the HR state does lead to $4q$
vacuum sectors on the torus, yet what is actually required for constructing a 
QHE state is a correlator containing many {\em insertions} of fields in the 
chiral algebra, corresponding to the particles in the ground state. For the HR 
case, these correlators are the ground states found above, and one set of $q$ 
sectors has turned out to contain two conformal blocks. However one chooses to
interpret this in CFT terms (it is probably related to other oddities of the
CFT for the HR state \cite{milr}), the existence of $4q$ sectors in the 
underlying theory is not in doubt.

Finally, we turn to the 331 state. Since the pole in $f_{\rm elliptic}$ is
expected to be first order, we try 
\begin{equation}
f_{{\rm elliptic},a}=\det\left[\frac{\vartheta_a((z_i^\uparrow-
z_j^\downarrow)/L_1|\tau)}
{\vartheta_1((z_i^\uparrow-z_j^\downarrow)/L_1|\tau)}\right].
\end{equation}
For $a=2$, $3$, $4$, these lead to nonvanishing, zero-energy states for the
pseudopotential Hamiltonian. We expect, however, that there are $4q$ states
altogether, based on the general structure of this abelian quantum Hall state
(see, e.g.\ Ref.\ \cite{milr}), and we expect the remaining $f_{\rm elliptic}$ 
to be $++$. Clearly the natural choice $\det \vartheta_1/\vartheta_1$ is
nonvanishing only for $N=2$. {}From the theory of elliptic functions in one
variable, the constant is the only elliptic function with at most one simple
pole and these boundary conditions, so we have exhausted the possibilities of
this structure. However, as pointed out above, $f_{\rm elliptic}$ need not be
an antisymmetrized product of elliptic functions in $(z_i-z_j)/L_1$ that each
satisfy the boundary condition; only $f_{\rm elliptic}$ itself must have this
property, so we should broaden our search.

The correct solution can be obtained, no doubt, in various ways. One way is to
use the expectation that $f_{\rm elliptic}$ is a conformal block for a Dirac
fermi field on the torus, with $++$ boundary condition in the case of interest
(the other functions $f_{{\rm elliptic},a}$, $a=2$, $3$, $4$, can be viewed in
exactly this way). This conformal block is known to exist, and could also be
obtained by bosonization. We are indebted to Greg Moore who obtained the 
following formula (for $N=4$) for this function, at our request, by a limiting
procedure of first considering the conformal block for the fermi field with the
general boundary condition, that is twisted by $e^{i\psi_1}$,
$e^{i\psi_2}$, and taking the limit $\psi_1$, $\psi_2\rightarrow 0$. The result
is (up to factors independent of the $z_i$'s)
\begin{equation}
f_{{\rm elliptic},1}=\frac{1}{(N/2-1)!}
\sum_{\stackrel{\scriptstyle\sigma\in S_{N/2}}{\rho\in
S_{N/2}}}{\rm sgn}\,\sigma \,{\rm sgn}\,\rho    
\prod_{r=1}^{N/2-1}
\frac{\vartheta_1'((z_{\sigma(1+r)}^\uparrow
-z_{\rho(1+r)}^\downarrow)/L_1|\tau)}
{\vartheta_1((z_{\sigma(1+r)}^\uparrow
-z_{\rho(1+r)}^\downarrow)/L_1|\tau)},
\label{331first}
\end{equation}      
where $\vartheta_1'(z|\tau)=d\vartheta_1(z|\tau)/dz$. Notice that, like the
fifth function for the HR state, there are two unpaired fermions of opposite
spins, occupying the constant single-particle state that is allowed by the
boundary conditions. For $N=2$, this is all that remains, and this function was
already noted above. For $N>2$ this function contains
$\vartheta_1'/\vartheta_1$, which is not strictly periodic (since no such
functions exist for $++$ boundary conditions), but obeys
\begin{equation}
\frac{\vartheta_1'(z+\tau|\tau)}{\vartheta_1(z+\tau|\tau)} 
=\frac{\vartheta_1'(z|\tau)}{\vartheta_1(z|\tau)}-2\pi i,  
\end{equation}
and is invariant under $z\rightarrow z+1$. When any $z_i^\uparrow$
($z_j^\downarrow$) in $f_{{\rm elliptic},1}$ is translated by $L_1\tau$
($-L_1\tau$), the result is $f_{{\rm elliptic},1}(z_1^\uparrow,\ldots,
z_{N/2}^\downarrow)$ plus a term that vanishes because the constant $-2\pi i$ 
must be antisymmetrized with the constant $1$ that represents the missing row
and column in the determinant. Thus $f_{{\rm elliptic},1}$ is invariant, and we
have found the fourth set of $q$ zero-energy states. We note that ${\bf k}=0$
for these states, and the quantum numbers of all $4q$ states are distinct, so
these states are linearly independent. 

In this section we have implicitly assumed that the number of particles $N$ is
even. One may ask if there are ground states on the torus for $N$ odd also. 
Such states will have an odd number of unpaired fermions. As we have seen, this
is possible only in the $++$ (or ${\bf k}=0$) sector. For the Pfaffian, there is
no such sector, so there are no ground states for $N$ odd, except for $N=1$. 
For 331 the ${\bf k}=0$ states already include two unpaired fermions for 
$N$ even. For $N$ odd, there must be just one unpaired fermion, of either up 
or down spin, otherwise the state vanishes. But, on generalizing eq.\ 
(\ref{331first}), one finds that it no longer satisfies the boundary condition,
except, again, for $N=1$. Finally, for HR,
there is no problem constructing a zero-energy state with one unpaired fermion 
in the ${\bf k}=0$ sector. This gives a spin-1/2 doublet of ground
states for all odd $N$, and we have verified numerically that states with
these quantum numbers are the only zero-energy states for $N$ odd. 
If these ground state wave functions are again interpreted as CFT correlators, 
then they imply that in this $++$ sector, there are nonzero correlators 
containing an odd number of insertions of the Fermi field. 

Greiter et al.\ \cite{gww} also found formulas for two quasiholes of the
Pfaffian state on the torus. There should be no great difficulty in extending
the results of the present paper to include any number of quasiholes on the
torus for any of the paired states we have considered.

\section{Effect of Zeeman term, tunneling, and other perturbations}
\label{tunneling}

In this section we address the question of what happens to the degeneracies of
the quasihole states and of the ground states on the torus when the Hamiltonian
is varied {}from the exactly-soluble (for the zero-energy states) forms we have 
considered up to now. First we consider the effect of the Zeeman term $-h\sum_i
\sigma_{z,i}$ on the HR state; here $h>0$ is the magnetic field $g\mu_B B$, 
and $\sigma_{z,i}$ is the $z$ component of the spin operator for the
$i$th particle, which can be represented by the usual Pauli matrices. The
hollow-core Hamiltonian is spin-rotation invariant, and its eigenstates are
also eigenstates for the total spin. Hence the Zeeman term simply has the
effect of splitting the multiplets of spin states. For the quasihole states,
this means that the degeneracy is partially resolved, independent of the
locations of the quasiholes. The lowest energy states
are then those with the largest number of $\uparrow$ unpaired fermions, and 
the lowest number of $\downarrow$. Since $F_\sigma\leq n-1$, and $F$ and $N$
have the same parity, these states contain $F_\uparrow=n-1$ and
$F_\downarrow=0$ when $N$ and $n-1$ are of the same parity, and either 
$F_\uparrow=n-1$, $F_\downarrow=1$ or $F_\uparrow=n-2$, $F_\downarrow=0$ when
$N$ and $n-1$ are of opposite parity. For fixed $w$'s these states clearly have
a residual degeneracy $1$ in the first case, $2(n-1)$ in the second. (The total
spin of these lowest-energy states, which is alternately $S_z=(n-1)/2$ or 
$S_z=(n-2)/2$, agrees with the results in Ref.\ \cite{haldrez}.) 
In the first case in particular, the lifting of the degeneracy implies that 
adiabatic exchange of the quasiholes can produce only a phase factor, and so 
the statistics is abelian. 

We also note here that for the edge states \cite{milr}, the gapless spectrum 
of spin-carrying fermion excitations at the edge implies
that the spin susceptibility of the edge is nonzero, and for finite $h$ there
will be some $\uparrow$ fermions present in the ground state, so there is a
magnetization at the edge, or one could say the edge is reconstructed in this
way; a confining potential is necessary to stabilise this effect. 
If one thinks of the $\uparrow$ fermions and their antiparticles, the 
$\downarrow$ fermions, as particles and holes of a chiral Fermi-Dirac sea, then
the reconstructed state just corresponds to shifting the Fermi energy of the
sea. There is thus still a gapless branch of fermion excitations for 
both spin $\uparrow$ and $\downarrow$ at the edge. 
Further, the degeneracy of $5q$ on the torus is not split by the Zeeman term,
because all the states are singlets.

It is clear that similar effects may be expected for any abelian or nonabelian
statistics state, when there is a symmetry present, particularly a continuous
symmetry: A symmetry-breaking perturbation may break the degeneracies and leave
some kind of abelian statistics behavior. This does not, however, necessarily 
mean that nonabelian statistics is unstable against {\em any} perturbation, nor
does it mean that the abelian statistics obtained is that of some simpler
abelian state, such as a Laughlin state of charge-2 bosonic pairs. We also 
note that, for either the Pfaffian or HR state, when there are only two 
quasiholes, there are no degeneracies for fixed quasihole positions on the 
sphere (so there is nothing to split), yet the 
expectation is that the Berry phase obtained on adiabatically exchanging the 
two is not that in the Laughlin state of charge-2 bosonic pairs \cite{mr}, at
least if the quasiholes are given by the wave functions studied in this paper. 
In fact, while the breaking of degeneracies does strictly mean that nonabelian
statistics does not occur in adiabatic transport of quasiparticles, other
associated properties, such as the spectrum of edge states, are, as we have
seen, not affected (in general, a splitting of the velocities might occur due
to a symmetry-breaking perturbation.) Quite similar phenomena are known in the 
hierarchy states, which have abelian statistics, where, for example, SU($N$) 
symmetry appears in the certain states \cite{read90}, but the symmetry 
(or related degeneracies) is presumably broken by the Hamiltonian, both for 
the bulk quasiparticle states and for the edge spectrum. For this reason, 
we propose that such effects do not really represent a change of universality 
class in the nonabelian systems either. 

We now discuss the effect of tunneling between the layers on the 331 states;
mathematically this is the same as a Zeeman-like term 
$-t \sum_i\sigma_{x,i}$ (see earlier discussions in \cite{halpnewport,ho}). 
This term can be diagonalized by using a basis of
$\sigma_x$ eigenstates for each particle, which we will label {\it e} and
{\it o} ( for ``even'', ``odd''), 
given by $e=(\uparrow+\downarrow)/\sqrt{2}$, 
$o=(\uparrow-\downarrow)/\sqrt{2}$, which have eigenvalues $+1$, $-1$ 
respectively under $\sigma_{x,i}$. In the literature, these states have often 
been denoted ``symmetric'' and ``antisymmetric'', respectively.   
Unlike the HR case, the pseudopotential Hamiltonian for which 331 is exact is
not spin-rotation invariant, and the energy eigenstates are not eigenstates of
$S_x=\frac{1}{2}\sum_i\sigma_{x,i}$. Thus the tunneling is a symmetry-breaking
perturbation, which breaks the conservation of $S_z$. The effect of the 
tunneling term is to modify the states, not merely to split their energies. 
Nonetheless, when $t>0$ is small, we may try to use degenerate perturbation 
theory to understand its effect, which means diagonalising the tunneling term 
in the subspace of the states that have zero energy when $t=0$; this would 
give the exact results at first order in $t$. We are not able to carry this out
analytically in general, because we do not have the matrix elements of the
perturbation among these states. We may expect, however, that the degeneracies
would be at least partially lifted, in a similar way as for the HR state with
Zeeman, by the following argument. 
In spin space, the pairs in the 331 state take the form
$\uparrow_i\downarrow_j+\downarrow_i\uparrow_j = e_ie_j-o_io_j$, using an
obvious notation for the spinors for the $i$th, $j$th particles in a term in
which these form a pair. Halperin \cite{halpnewport} has
argued that the effect of positive tunneling $t$ is to cause a change in the 
331 ground state, which within a trial-wave-function
description causes the amount of $oo$ in the pairs to decrease. 
Ho \cite{ho} has proposed that this occurs at first order in $t$. If we write 
the unpaired fermions in our zero-energy states also in the $e$-$o$ basis, and
neglect the effect of $t$ on the paired part of the state, its effect would be
to split the energies of the $e$ and $o$ unpaired fermions, exactly as for the 
$\uparrow$ and $\downarrow$ fermions of the HR state, and again with the
effect of removing most of the degeneracy. For the ground states on the torus, 
we first note that the states with ${\bf k}\neq0$ are even under layer
exchange, which has the effect of multiplying by $\prod_i \sigma_{x,i}$, while
the states at ${\bf k}=0$ are odd (incidentally, this agrees with numerical
findings \cite{haldpriv}). The broken pair in those states contains one
$e$ and one $o$ fermion. Application of the same naive argument as for the
quasihole states then suggests that the $4q$ degeneracy is split to $3q$ by
$t>0$. However, an accurate calculation should include the modification of the
rest of the state, and the splitting might disappear in the
thermodynamic limit. 

If we consider arbitrary small changes in the Hamiltonian {}from the special
forms considered in previous sections, then physical arguments like those for
the Zeeman and tunneling terms suggest that, as the degeneracy arises {}from
breaking pairs and putting the fermions into the zero-mode functions, there is
no obvious reason why it should not be broken in general. For the bound
states in the gap in the vortex cores in conventional BCS superfluids, such
excited many-particle states have positive energies. However, such
arguments may just be too naive, because of the modification of 
the ground state wave function.

In an interesting paper, Ho \cite{ho} has given a more general interpolation 
between the 331 and Pfaffian ground states by varying the spin states of the
pairs. Within a paired trial wave function description, the effect of tunneling
is presumably to reduce the amount of $oo$ in the spin part of the 
wave function of a pair \cite{halpnewport}. If this is done without changing 
the spatial factor $(z_i-z_j)^{-1}$, then when the pairs
are purely $ee$, the ground state is precisely the Pfaffian state. 
Ho claims that this somehow contradicts the ``topological'' arguments that
assert that one cannot go continuously between these distinct ground states.
However, this is really a misstatement; one can {\em always} interpolate 
between any
two state vectors in the same Hilbert space. The real question is whether the
{\em properties} of the states, like those considered in this paper, can be 
continuously connected. If one wishes to exhibit
a breakdown of the ``topological'' arguments, then it is necessary 
to show that the interpolation occurs without any phase transition, that is 
without any energy gap for local excitations going to zero (which would be a 
second order transition) or any level crossing of ground states (which would 
be a first order transition). Ho proposed a family of Hamiltonians for each of 
which his corresponding wave function is a zero-energy ground state, but did 
not show that the energy gap is maintained throughout the interpolation. We 
will now examine this. We will show that at the point where Ho's ground state 
is the Pfaffian, there is an enormous degeneracy of other zero-energy states 
for his Hamiltonian, implying that the energy gap collapses to zero at this 
point, and the system is not incompressible. We will show that, up to that 
point, the degeneracies of the zero-energy eigenstates of his model are the 
same as those of the pseudopotential Hamiltonian for the 331 state, considered 
earlier. We will then consider modifications of his model that remove the
pathology.

Following Ho, we now fix $q=2$, so the particles are fermions, and
the filling factor is $\nu=1/2$ (as usual, other cases are similar). 
To understand Ho's description, first recall 
that for spin 1/2 fermions in the LLL, Fermi statistics implies that if the 
particles have even relative angular momentum, $m=0$, $2$, \ldots, then they 
must be in an antisymmetric spin state, which can only be a singlet, while if 
they are in an odd relative angular momentum state, $m=1$, $3$, \ldots, 
then they can only be in a symmetric spin state, which must be a triplet. We
emphasize that these statements remain true in the presence of any number of
other particles. If the two particles have total spin $1$, then there is a 
three-dimensional complex vector space of spin states, and the spin state can be
described exactly by a nonzero complex 3-component vector $\bf d$, of which the
magnitude and phase are irrelevant to the state. We will not need the detailed
definition of $\bf d$, which can be found in Ho \cite{ho} or in references on
superfluid $^3$He (see e.g.\ Ref.\ \cite{vollhardt}). We note only that, 
for the spin state $\uparrow_i\downarrow_j+\downarrow_i\uparrow_j = 
e_ie_j-o_io_j$ (as in the pairs in the 331 state), 
${\bf d}\propto \hat{\bf x}_z$, while for $e_ie_j$ (as in the pairs in the
Pfaffian), ${\bf d}\propto (\hat{\bf x}_z-i \hat{\bf x}_y)/\sqrt{2}$ (here 
$\hat{\bf x}_z$ denotes a unit vector in spin space in the $z$ direction, etc.).
Notice that the transformation {}from one state to the other is not simply a
rotation. Ho's Hamiltonian is a pseudopotential Hamiltonian 
\begin{equation}
H_{\rm Ho}({\bf d})=\sum_{i<j}\left[V_0P_{ij}(N_\phi,S=0)+V_1P_{ij}(N_\phi-1,
\perp{\bf d})\right]
\label{HopseudH}
\end{equation}
which gives positive energy to any two 
particles with (i) relative angular momentum zero and spin zero, or (ii) 
relative angular momentum one and triplet spin state orthogonal to a chosen 
state specified by $\bf d$. For the choice of $\bf d$ that corresponds to 331, 
this reduces to the pseudopotential Hamiltonian (\ref{331pseudH}) used before; 
pairs may have 
zero energy and relative angular momentum one only if they are in the 
$\uparrow_i\downarrow_j+\downarrow_i\uparrow_j$ spin state. For the $\bf d$ 
vector that corresponds to the $ee$ pairs in the Pfaffian, it allows two 
particles to have relative angular momentum one at zero energy if they both 
have spin $e$, not if they are $eo+oe$ or $oo$. However, at this ${\bf d}$
vector, which we call the Pfaffian point, {\em all states of the electrons in 
which all spins are $e$ are zero-energy eigenstates of this Hamiltonian}. This 
follows because there are clearly no singlets in such states, so no relative 
angular momentum zero pairs either. This should be no surprise, since no 
convenient 
two-body Hamiltonian (like Ho's) giving the Pfaffian as ground state and a 
sensible spectrum is known. Since there are very many spin-aligned states at 
$N_\phi=2N-3$, Ho's Hamiltonian has a very large ground state degeneracy. 
There might, of course, also be degenerate zero-energy states in which the 
spins are not all $e$. We will fully analyse this degeneracy below.

For ${\bf d}\not\propto(\hat{\bf x}_z-i \hat{\bf x}_y)/\sqrt{2}$, the
degeneracies of Ho's ground state, and of quasihole states on the sphere and
ground states on the torus, coincide with what was found earlier for the case
of the 331 state. To construct these zero-energy states for general $\bf d$, we
use a more precise notation for the wave functions that includes the spin 
states.
We label the particles $i=1$, \ldots, $N$, and use $\uparrow_i$, 
$\downarrow_i$ for spinors which are eigenstates of $\sigma_{z,i}$; the 
wave function is now in a tensor-product space of spatial wave functions and 
spinors, and must be antisymmetric under simultaneous exchange of coordinates 
and spinors of two particles. For example, the ground state on the sphere is
\begin{equation}
\tilde{\Psi}_{\bf d}=\sum_{\sigma\in S_N} {\rm sgn}\,\sigma
\prod_{k=1}^{N/2}\frac{\chi_{\sigma(2k-1),\sigma(2k)}({\bf d})}
{z_{\sigma(2k-1)}-z_{\sigma(2k)}}\prod_{i<j}(z_i-z_j)^q
\label{hostate}
\end{equation}
(which is actually a Pfaffian), where $\chi_{i,j}({\bf d})$ is the spin state of
two particles $i$, $j$ that corresponds to $\bf d$, and the product of these
factors is actually a tensor product, so that $\tilde{\Psi}_{\bf d}$ is
multi-spinor function of the coordinates. This clearly has zero
energy. 
We will call it the Ho state, as it appears in Ref.\ \cite{ho} (in a
different notation). 
In the presence of quasihole factors $\Phi$ in the 
pairing factors, one may construct zero-energy states with unpaired fermions, 
as for the 331 state, but with the pairs taking the same form as in eq.\ 
(\ref{hostate}). The unpaired fermions can be in either spin state. Even though 
the ground state (\ref{hostate}) does not have definite total $S_z$, except 
in the 331 special case, the counting of zero-energy quasihole states proceeds 
just as for the 331 state, and the results are identical to those in Sec.\ 
\ref{331}. Similar results are found for the edge states, which are in 
one-to-one correspondence with those in \cite{milr}, and for ground states 
on the torus, which are like those in Sec.\ \ref{torus}. We expect that 
these zero-energy states are the complete set, except in the Pfaffian limit. 
This is consistent with the continuity of the spectrum as a function of 
$\bf d$ {\em in a finite size system}, which
requires that the energy levels found in the 331 state 
must change continuously with the Hamiltonian. We expect that the larger
degeneracy appears only at the Pfaffian point, because there the pairs have the
special property of being composed of one spin ($e$) only, unlike the general
states $\chi({\bf d})$. This implies that the energy of some excited states 
decreases as this point is approached, so the gap goes to zero. Therefore Ho's
Hamiltonian is pathological at the Pfaffian point. 

In spite of this pathology at the Pfaffian point, it is still of interest, 
given that the degeneracies of the quasiholes and torus ground states 
are the same in Ho's model (except for the Pfaffian point) as in the 331 
state, to ask whether the statistics properties are the same. In the 331
state, the structure of these properties is described in terms of a U(1)
[or SO(2)] quantum number, which in the case of 331 is $S_z$. 
Also, excitations with opposite values of this quantum number are degenerate,
because of the symmetry operation of interchanging the layers, which generates
a ${\bf Z}_2$ group. These two symmetries combine to
make up the symmetry group O($2$), which is the semi-direct product of 
${\bf Z}_2$ and SO($2$). Ho's Hamiltonian breaks the conservation of 
$S_z$, but not the ${\bf Z}_2$ symmetry, if we consider only the family 
${\bf d}=\hat{\bf x}_z\cos\theta -i\hat{\bf x}_y\sin\theta$, 
$0\leq\theta\leq\pi/4$ as proposed by Ho; 
these correspond to the spin state $\cos(\theta-\pi/4)e_ie_j
+\sin(\theta-\pi/4)o_io_j$ for the pairs, suggested by Halperin 
\cite{halpnewport}, and $\theta=0$ is the 331 case, $\theta=\pi/4$ is the
Pfaffian. 
Consequently, one might think the U(1) quantum number is lost. However, 
the degeneracies of the quasiholes and torus ground 
states are consistent with the presence of this quantum number, which would be 
a ``hidden'' U(1) symmetry. One might expect this to be, in some sense,
the symmetry of rotation in spin space about the axis of $\bf d$, however,
since $\bf d$ is complex we must be careful. Under rotations of
spin space, $\bf d$ is rotated by the action of a (real) orthogonal $3\times3$ 
matrix in O(3). In general, there is no rotation that leaves $\bf d$ invariant 
(up to multiplication by a phase), except when $\bf d$ is of certain special 
forms of which the $\bf d$ vectors for 331 and the Pfaffian happen to be 
examples. Remarkably, even though O(2) symmetry is broken by Ho's Hamiltonian, 
it seems to be reappearing in the low-energy properties. Indeed, in terms of a 
conformal field theory (CFT) description of the edge states on the cylinder, 
which gives detailed information about the structure of the states \cite{milr},
the low-energy edge states obtained as zero energy eigenstates of Ho's model 
have the same structure as in \cite{milr} for all $\bf d$. If we assume that a 
CFT description must involve a U(1) theory for charge, together with some 
unitary $c=1$ theory, combined by the ${\bf Z}_2$ orbifold construction of 
Ref.\ \cite{milr}, then the theory described in that reference seems to 
be the only possibility. However, we should be cautious about concluding on the
basis of these observations that 
the universality class of Ho's model is the same as that of the 331 state, 
except at the Pfaffian point. We will see below an example in which the full
properties of this class do not emerge, and the degeneracy can be broken by a
perturbation even in the thermodynamic limit. This example emerges {}from 
further analysis of the Pfaffian point of Ho's model, to which we now turn.

At the Pfaffian point of Ho's model, in the $q=2$ case, it is a two-body 
projection-operator Hamiltonian which gives positive energy to any two 
particles which either have opposite spin and relative angular momentum zero, 
or if one or both of them is $o$ and they have relative angular momentum one. 
It is known in general how to find the zero-energy states of such 
pseudopotential Hamiltonians; this was already discussed at the beginning of 
Sec.\ \ref{331}. To be zero-energy eigenstates, wave functions must contain 
the $mm'n$ wave function, eq.\ (\ref{mmnground}), as a factor, in which for the 
particular class (but for general $q$) considered here, we have $m=q-1$, 
$m'=q+1$, $n=q$, 
\begin{equation}
\tilde{\Psi}_{q-1,q+1,q}(z_1^e,\ldots,z_{N_e}^e,
z_1^o,\ldots,z_{N_o}^o)=
\prod_{i<j} (z_i^e-z_j^e)^{q-1}
\prod_{k<l} (z_k^o-z_l^o)^{q+1} 
\prod_{rs} (z_r^e-z_s^o)^{q}. 
\label{pfptground}
\end{equation}
For functions on the sphere, the number of flux seen by $e$
and by $o$ particles must be the same, but this is not the case for the
function (\ref{pfptground}) as it stands, unless $N_o$ is zero. If $N_o>0$, 
it is necessary to multiply the function by additional factors, and the space
of these factors may be parametrised by viewing them as quasihole factors 
$U_\sigma(w^\sigma)=\prod_i(z_i^\sigma-w^\sigma)$, where $\sigma=e$ or $o$, 
acting on particles of either spin. On multiplying in $n_e$ factors of $U_e$,
$n_o$ factors of $U_o$, one finds for the flux seen by $e$ and $o$
respectively,
\begin{eqnarray}
N_\phi&=&(q-1)(N_e-1)+qN_o+n_e\\
  &=&(q+1)(N_o-1)+qN_e+n_o
\end{eqnarray}
since the flux must be the same for both; the second line, which is the flux
seen by the $o$ particles, applies only if $N_o>0$. 

We wish to analyse the situation $N_\phi\geq q(N-1)-1$, that corresponds to the
Pfaffian ground state or the same plus quasiholes. If we write 
\begin{equation}
\Delta N_\phi=N_\phi-[q(N-1)-1],
\end{equation}
which was denoted $n$ in Sec.\ \ref{pfaffian}, then we find the $q$-independent
equations
\begin{eqnarray}
n_e&=&N_e-2+\Delta N_\phi,\\
n_o+N_o&=&\Delta N_\phi,
\end{eqnarray}
where again the second equation does not apply if $N_o=0$. Now we see that if
$\Delta N_\phi\leq0$, we must have $N_o=0$ since $n_o\geq0$. So in this region,
all zero-energy states contain only $e$ particles. In particular, 
this includes the Pfaffian state and all states degenerate with it at 
$\Delta N_\phi=0$; for $q=1$, this space of states includes all states in which
all the bosons have spin $e$, and the same is true for the fermions at $q=2$, 
as remarked above, since antisymmetry requires all states to include the factor 
$\prod_{i<j}(z_i^e-z_j^e)$ as a factor. The nondegenerate ground state of the 
model occurs at $N_\phi=(q-1)(N-1)$ (i.e.\ at $\Delta N_\phi=-(N-1)+1$), 
and for $q=2$ is the LLL filled with $e$ 
particles. For $\Delta N_\phi>0$, 
the maximum number of $o$ particles possible in the zero energy states is 
$N_o\leq \Delta N_\phi$. 
The numbers of zero-energy states, and their angular-momentum decomposition, 
for each $N$, $N_\phi$ for this model can now be obtained, as in other
cases (see especially the case of the Laughlin state, and the 331 state 
in the bosonic language). 

Numerical study of Ho's model confirms the above discussion. Figures
\ref{fig:ho0qh} and \ref{fig:ho0qhenl} show representative spectra with no 
quasiholes, for
various values of $\theta$, including the 331 point $\theta=0$, and for $N=6$,
$q=2$. We chose $V_0=1$, $V_1=0.5$.
For $\bf d$ not at the Pfaffian point, the zero-energy ground state at
$N_\phi=q(N-1)-1$, is nondegenerate. As the Pfaffian point is approached, 
a set of states approaches zero energy, and at the Pfaffian 
point, the degeneracies of the zero-energy states are exactly those of the 
$mm'n$ system that we have just analyzed. 
Already at
$\theta=\pi/8$, one can identify most of the states that reach zero energy at
the Pfaffian point. Calculations not shown in the Figures also confirm that 
states with quasiholes have the degeneracies, and the 
angular momenta, of those for the 331 state. We notice that,
in addition to the zero-energy states at the Pfaffian point, which are fully
explained by the above analysis, there are also some very low energy excited
states, for which we do not at present have a detailed explanation.

There is a simple way to remove the ``excess'' degeneracy at the Pfaffian
point, without destroying the 331-like behavior of zero-energy states
elsewhere. The Pfaffian, with all particles $e$, would be selected by the
three-body Hamiltonian used earlier in the spinless case, if it acted on the
$e$ particles. We also observe, {}from the structure of the wave functions
(\ref{hostate}), that in these wave functions no three particles 
have total angular momentum $3N_\phi/2-3(q-1)$. 
Therefore, the three-body operator which projects each set of three particles 
onto angular momentum  $3N_\phi/2-3(q-1)$ and spin 3/2 (or its analog on the
torus) annihilates the 
Ho states, and the quasihole states and ground states on the torus derived 
{}from it {\em for all ${\bf d}$} (the quantum numbers are most easily derived 
by considering spin-1/2 
bosons for $q=1$; in these Ho states, three bosons are never found at the same 
point---if they were they would be in a symmetric spin state). If it is added
to the Ho Hamiltonian, all the degeneracies of zero-energy states will be
maintained, away {}from the Pfaffian point. At the Pfaffian point, the total
Hamiltonian now selects, at $N_\phi=q(N-1)-1$, the Pfaffian ground state
as the unique zero-energy state, in which all spins are $e$. For smaller
$N_\phi$ there are no zero-energy states, and for larger $N_\phi$ the zero
energy states are just those of the Ho form (\ref{hostate}) and its
generalizations. In these states, the paired fermions are all $e$, but the
unpaired ones can be either $e$ or $o$. Consequently, the degeneracies are again
those of the 331 state, {\em not those of the Pfaffian}. This was of course 
inevitable by continuity, given that no states now come down to zero energy at 
the Pfaffian point. Numerical spectra confirm these predictions, as shown in 
Fig.\ \ref{fig:ho+3bod0qh} for $0$ quasiholes, and in Fig.\
\ref{fig:ho+3bod4qh} for $4$ quasiholes. In these figures, the coefficient of 
the three-body projection operator is 1.

We have now arrived at a Hamiltonian which has the degeneracies of the 331
state at all $\bf d$ vectors. Yet the Ho model was supposed to represent the
effect of tunneling, which should raise the $o$ particles to high energy, and
it was expected that the ground state for strong tunneling would be the 
Pfaffian. While the ground state we find at the Pfaffian point is the Pfaffian,
the problem is that unpaired particles of either spin can be in the zero modes. 
But this can now be easily (and exactly) remedied by adding to the Hamiltonian 
the tunneling term $-t\sum_i \sigma_{x,i}$. Since the tunneling term is 
diagonal in the $e$-$o$ basis we are using, it simply 
splits the states we have found. Clearly, all states containing only $e$ 
particles now have energy $-Nt$, and these have the degeneracies of the 
Pfaffian state for any number of quasiholes, or for edge states, or on the 
torus. States containing $o$ particles are higher in energy. If we consider 
the same model, as a function of $\bf d$, and let $t$ 
vary with $\bf d$, such that $t$ is zero at the 331 point and of order one at 
the Pfaffian point, then we can say that its lowest-energy states are known 
exactly at the 331 and Pfaffian points, but unfortunately not in-between.  
There must be one or more phase transitions as the Hamiltonian is varied 
between these limits. Notice that there is a surpising effect at the 
Pfaffian point, which has edge states including unpaired fermions of 
either spin when $t$ is zero, but the $o$ states obtain a gap when $t>0$. 

When studying more general Hamiltonians that include tunneling, 
Ho's state would in general be a better choice of trial state than the 
original 331 state, and $\bf d$ can be used as a variational parameter. 
We wish to emphasize that a large overlap of the ground state of a Hamiltonian
with a particular trial state, say, the 331 state (as in \cite{he}) 
does not prove that the state is in the corresponding (say, 331) universality 
class; to show that, its properties must be calculated, and the 
thermodynamic limit taken, and this is a more difficult task numerically. 
Conversely, a low overlap would not prove it is not in that class.

Our discussion still leaves the question of
whether for large enough $t$ there is a transition to the Pfaffian universality
class, or to some other class. Because of the similarity in ground state
structure of the 331 and Pfaffian states emphasized in Ref.\
\cite{gww,halpnewport,ho}, a second order transition seems to be a possibility.
Another possibility is that the Halperin class of paired states
is involved, those for which a trial state can be constructed by first pairing
the particles into charge-2 bosons, then forming a $\nu_b=1/4q=1/8$ state of the
bosons \cite{halp83,gww}. This is clearly another abelian quantum Hall state,
for which the edge excitations would consist only of the U(1) 
density-fluctuation chiral scalar boson modes, in different charge sectors; it
would lack the gapless fermion excitations characteristic of the paired states
discussed here \cite{milr}. 

While we cannot rule out the Halperin type of paired state, and all of these
states might describe the universality classes of various Hamiltonians, even 
within the LLL, we can point out an experimental signature that will 
distinguish the former {}from the others. 
First we recall that, for the Laughlin states, the Luttinger liquid at the edge
leads to the local density of states $N(\omega)\sim |\omega|^{q-1}$ for filling
factor $1/q$ (see Ref.\ \cite{wenrev} for a review). The corresponding
tunneling differential conductance for a point contact at zero temperature 
will be $dI/dV\sim V^{q-1}$. The exponents are simply related to the scaling 
dimension $\Delta$ of the operator that creates a single electron in the 
low energy theory of the edge, as $dI/dV\sim V^{2\Delta-1}$, and also 
to the statistics $\theta$ of the excitation by $\theta/2\pi\equiv\Delta$ 
(mod $1$). 
In the theory for the Laughlin states, the charge $n/q$ operators at a 
single edge have dimensions $n^2/2q$ ($n$ must be an 
integer); the non-integral charge operators must be combined with similar
operators at the other edge (see Refs.\ \cite{wenrev} or \cite{milr}). Thus
$\Delta=q/2$ for the electron.
In the charge-$2$ boson system, the charged operators that create charges 
at the edge within the low-energy theory 
(subject to some straightforward conditions discussed in Ref.\ \cite{milr}) 
have scaling dimension $n^2/8q$ when they create charge $\pm n/2q$ 
(in electron units) at the edge at filling factor $1/q$, again for $n$
integral. As in the old argument of Tao and Wu \cite{taowu}, 
in either theory the operator of charge $1$ has $\theta=2\pi\Delta=\pi q$, 
which is Bose statistics for $q$ even, and cannot represent an electron
(similarly, for $q$ odd, it cannot represent a charge-$1$ boson). In addition, 
such an
operator cannot be used at a single edge in the charge-2 boson theory, but must
be combined with another charged operator at the other edge, or with some 
other operator at the same edge. Only operators with even-integer charge can
act on one edge.
Therefore, at low energies, (that is, bias voltages), tunneling into the edge 
{}from outside will be impossible at these filling factors $\nu=1/q$ for this 
charge-2 boson universality class. To make an electron on one edge, 
the charged operator must be combined with an operator making an unpaired 
fermion (or BCS quasiparticle), which in the present case would exist as 
excitations, but would have a finite energy gap, even at the edge; thus the 
tunneling current will be zero below a threshold voltage. On the other hand, 
in the paired states like the Pfaffian, HR and 331 states, the fermion
excitations are gapless at the edge, and a nonzero tunneling current with a
power-law dependence at small bias voltage should be observable (the power law
can be calculated {}from the theories in Ref.\ \cite{milr}; $\Delta$ has an
extra contribution {}from the fermions, to yield 
$V^q$, for both the Pfaffian and 331 cases, and $V^{q+1}$ for HR). 
These arguments are for leading order in the tunneling. It is possible that 
some sort of higher-order tunneling process could transfer two electrons into 
the edge, through a virtual transition to one or more higher-energy states; 
as this does not require any fermions to be created except virtually, 
this would give a current at arbitrarily low bias voltage. However, 
the exponent would be related to the scaling dimension (and statistics) 
for the charge-2 operator in the low-energy edge field theory of the 
charge-2 boson state, and so the power law would be $V^{4q-1}$; thus 
for $q=2$ the current would be much lower than for Pfaffian, 331 and HR states, 
and still clearly distinguishable. 
For another discussion of the experimental consequences of the 
Halperin-type charge-2 boson Laughlin state, in connection
with even-integer filling factors, see Ref.\ \cite{tikkiv}; our formulas for 
the charge-2 boson universality class apply for the conductance 
in this case also, with $q=1/2$, to give $dI/dV\sim V$.

\section{Conclusion}
\label{conclusion}

In summary, we have obtained a full description of the quasihole states of
several paired FQHE states, for the Hamiltonians for which the exact ground
states are known. The degeneracies found in the Pfaffian and HR cases are as
required for nonabelian statistics. For the 331 states, the statistics are
abelian, and the degeneracies are due to the layer index. Ho's model was found
to be pathological at its Pfaffian point, but the pathology was removed by
adding a three-body term to Ho's Hamiltonian. With tunneling added also, the
Pfaffian state was recovered, but the model was no longer exactly soluble 
for the low-energy states at intermediate points in the parameter space 
between the 331 and Pfaffian points. In Appendix \ref{perm}, the permanent 
state was also considered, which is another candidate for nonabelian 
statistics, but should probably be rejected because of its proximity to a 
ferromagnetic ground state, and its correspondingly gapless nature. It remains 
to be seen whether the other nonabelian states, though not apparently close to 
an obviously gapless state, are in fact stable against small generic 
perturbations. This is an important outstanding issue to which we hope to 
return elsewhere. It probably requires an analytical, field-theoretic 
technique to settle it in general, which should be a theory that describes the 
paired condensate, and not just a Chern-Simons theory of the low-energy sector 
containing nonabelian statistics. In the meantime, we have pointed out in 
Sec.\ \ref{tunneling} how the different paired states, the Halperin-type state 
of charge-2 bosons, and the Pfaffian and HR types with nonabelian statistics, 
can be distinguished in a point-contact tunneling experiment. As for an actual 
demonstration that adiabatic transport of quasiparticles does produce 
nonabelian statistics in some systems, that also will have to be left for 
treatment elsewhere.

\acknowledgements
We thank G.~Moore, M.~Greiter, D.~Haldane, B.~Halperin, A.~Kol, and 
M.~Milovanovi\'{c} for helpful conversations. 
N.R. acknowledges support {}from NSF grant DMR-9157484, and {}from the A.P. 
Sloan Foundation. E.R. was supported by NSF-DMR-9420560.

\appendix

\section{Elliptic functions}
\label{theta}

In general, theta functions with characteristics are defined as 
\begin{equation}
\vartheta \left[\begin{array}{c}
a\\
b \end{array}\right] (z|\tau)=\sum_{n}
e^{i\pi\tau(n+a)^2 +2\pi i(n+a)(z+b)}
\end{equation}
where the $n$ sum is over all the integers and $a$ and $b$ are real.
{}From the definition
we obtain:
\begin{eqnarray}
\vartheta \left[\begin{array}{c}
a\\
b \end{array}\right] (z+1|\tau)&=&
e^{2\pi i a}\vartheta \left[\begin{array}{c}
a\\
b \end{array}\right] (z|\tau)
\nonumber\\
\vartheta \left[\begin{array}{c}
a\\
b \end{array}\right] (z+\tau|\tau)&=&
e^{-i\pi\tau-2\pi i(z+b)}\vartheta \left[\begin{array}{c}
a\\
b \end{array}\right](z|\tau)
\end{eqnarray}
Consequently we can restrict $a$, $b$ to lie between $0$ and $1$. 

The standard Jacobi theta functions \cite{copson} are 
particular cases of those above. There are four of them: $\vartheta_1(z|\tau)=
\vartheta\left[\begin{array}{c}
1/2\\
1/2 \end{array}\right]$, $\vartheta_2=\vartheta\left[\begin{array}{c}
1/2\\
0 \end{array}\right]$, $\vartheta_3=\vartheta\left[\begin{array}{c}
0\\
0 \end{array}\right]$, and
$\vartheta_4=\vartheta\left[\begin{array}{c}
0\\
1/2 \end{array}\right]$. Of these, $\vartheta_1$ is of particular importance,
since it is odd under $z\rightarrow -z$, and so has a zero at the origin. In
general, all  $\vartheta\left[\begin{array}{c}
a\\
b \end{array}\right]$ can be related to each other by shifts of $z$.

The identity used to rewrite the $F_{\rm cm}$ found in Ref.\ \cite{hr85} 
in the form given in Sec.\ \ref{torus} can be obtained, by shifts of $z$, 
{}from the simplest version
\begin{equation}
\prod_{r=1}^M \vartheta_3(z-r/M|\tau)=\vartheta_3(Mz+(M-1)/2|M\tau)
\eta^M(\tau)/\eta(M\tau),
\end{equation}
which is obtained by writing the theta functions on the left-hand side in 
the product form \cite{copson}
\begin{equation}
\vartheta_3(z|\tau)=\prod_{n=1}^\infty(1-e^{2\pi i\tau n})(1+
2e^{2\pi i\tau(2n-1)}
\cos2\pi z +e^{4\pi i\tau(2n-1)}) 
\end{equation}
and doing the $r$ product first. Here $\eta(\tau)$ is the
Dedekind $\eta$ function, 
\begin{equation}
\eta(\tau)=e^{i\pi \tau/12}\prod_{n=1}^{\infty}(1-e^{2\pi i\tau n}).
\label{eta}
\end{equation} 

The Weierstrass elliptic function can be defined by \cite{copson}
\begin{equation}
\wp(z|\tau)=\frac{1}{z^2}+{\sum_{m,n}}'\left\{\frac{1}{(z-m-n\tau)^2}
-\frac{1}{(m+n\tau)^2}\right\},
\end{equation}
where the prime indicates that $m=n=0$ is to be omitted {}from the sum. It can 
be shown \cite{copson} that $\wp$ is invariant under $z\rightarrow z+1$, 
$z\rightarrow z+\tau$.
To obtain functions that have a double pole at the origin, like the Weierstrass
function, but are antiperiodic, we may use the definitions
\begin{equation}
\wp_{a,b}(z|\tau)=\sum_{m,n}(-1)^{ma+nb}\frac{1}{(z-m-n\tau)^2}
\end{equation}
where $a$, $b$ are integers. $\wp_{a,b}$ depends on $a$ and $b$ only modulo 2. 
For $a$, $b$ both even, the sum is not convergent, which is why $\wp$ was not
defined this way. For the other three cases, the series converges, and it is
clear that the functions have the periodicity properties that we denoted $+-$,
$-+$, $--$ in the text. 

\section{Permanent state}
\label{perm}

In this Appendix, we will introduce a Hamiltonian for which a certain state
containing a permanent \cite{mr} (other such states can be found in Ref.\
\cite{ymg}), is the unique zero-energy eigenstate of maximum density, 
and summarise results for the quasihole and edge excitations, and for the 
ground states on the torus. We also describe a relation with fully 
spin-polarized states and their skyrmion excitations, and argue that the
permanent state is at a phase transition {}from ferromagnet to paramagnet. 

The permanent state is a spin-singlet ground state
of spin-$1/2$ fermions for $q$ odd, and of spin-$1/2$ bosons for $q$ even. 
It can be viewed as spin-singlet $p$-wave pairing of composite bosons \cite{mr}.
The Hamiltonian for the simplest case, $q=1$, is a three-body Hamiltonian, 
which penalises the closest possible approach of three spin-1/2 fermions. 
On the sphere, three particles are at their closest, consistent with 
Fermi statistics, when the total orbital angular momentum for the three is 
$3N_\phi/2-1$, and the total spin is $1/2$. Our three-body Hamiltonian is
therefore taken to be a positive number times the projection operator 
onto this multiplet of states for three particles, summed over all triples of
particles:
\begin{equation}
H=\sum_{i<j<k}VP_{ijk}(3N_\phi/2-1,1/2).
\label{perm3bodH}
\end{equation}
 We have verified numerically that this does produce a unique
many-particle state at zero energy at the $N_\phi$ value that corresponds to
the permanent state at $\nu=1$. As the permanent state is a zero-energy state
for this Hamiltonian by inspection (and is non-zero), it must be the state 
obtained numerically. As for the Pfaffian state in Sec.\ \ref{pfaffian}, 
a suitable Hamiltonian, consisting of a combination of spin-independent 
two-body projection operators onto angular momenta $N_\phi$, $N_\phi-1$, 
\ldots, $N_\phi-q+2$, and a certain three-body projection onto angular momentum 
$3N_\phi/2-3(q-1)-1$ and spin 1/2, can be constructed for which the 
generalizations of these states to $q>1$ are again the complete set 
of zero energy states.

In our much-abused notation, the permanent state is 
defined by the wave function \cite{mr}
\begin{equation}      
\tilde{\Psi}_{\rm perm}(z_{1}^{\uparrow},\ldots,z_{N/2}^{\uparrow},
z_1^{\downarrow},\ldots,z_{N/2}^{\downarrow})=
\sum_{\sigma\in S_{N/2}}
         \frac{1}{(z_1^\uparrow-z_{\sigma(1)}^\downarrow)
           \cdots(z_{N/2}^\uparrow-z_{\sigma(N/2)}^\downarrow)}
\prod_{i<j}(z_i-z_j)^q.
\label{permground}
\end{equation}           
The fact that it represents a singlet is most easily seen by viewing it as
singlet pairs of composite bosons of spin $1/2$. It is totally
antisymmetric for $q$ odd, symmetric for $q$ even. It resembles the 331
state, but with the sign of the permutation omitted. Such a summation over
permutations defines the permanent of a matrix, ${\rm perm}\,M=\sum_\sigma
\prod_i M_{i,\sigma(i)}$, in which $M$ is an $L\times L$ matrix with elements 
$M_{ij}$, and the sum is over all members $\sigma$ of $S_L$.

States with $2n$ quasiholes can be written down in the now-familiar manner:
\begin{eqnarray}
&&\frac{1}{(N_\uparrow-B_\uparrow)!}
\sum_{\stackrel{\scriptstyle\sigma\in S_{N_\uparrow}}{\rho\in
S_{N_\downarrow}}}
\prod_{k=1}^{B_\uparrow} (z_{\sigma(k)}^\uparrow)^{n_k}
 \prod_{l=1}^{B_\downarrow} (z_{\rho(l)}^\downarrow)^{m_l}
\prod_{r=1}^{N_\uparrow-B_\uparrow}
\frac{\Phi(z_{\sigma(B_\uparrow+r)}^\uparrow,
z_{\rho(B_\downarrow+r)}^\downarrow;w_1,\ldots,w_{2n})}
{(z_{\sigma(B_\uparrow+r)}^\uparrow -
z_{\rho(B_\downarrow+r)}^\downarrow)}\nonumber\\
&&\qquad\qquad\qquad\times\prod_{i<j}(z_i-z_j)^q,
\label{perm+2nqholes}
\end{eqnarray}              
in which $\Phi$ is as in Secs.\ II--IV, and $n_k$, $m_l$ must be $\leq n-1$. 
In these states the flux is $N_\phi=q(N-1)-1+n$.
It is clear that these are all
zero-energy states for the three-body interaction eq.\ (\ref{perm3bodH}) 
at $q=1$, and its generalization to $q>1$. The counting of
states is most similar to the 331 state, but because the unpaired particles are
here spin-$1/2$ bosons, the Pauli principle restriction on the number of
unpaired particles does not apply, and there is no upper limit, except that the
numer of unpaired particles cannot exceed the total number of particles. 
For $n\geq 1$, the number of states for fixed $w$'s, and a fixed number of 
broken pairs, is that for $B=B_\uparrow+B_\downarrow$ unpaired bosons, 
in $2n$ orbitals (including spin $\uparrow$ or $\downarrow$), which yields
\begin{equation}
\left(\begin{array}{c}
B+2n-1\\
B \end{array}\right). 
\end{equation}
Including the positional degeneracy of the quasiholes, and summing over $B$ as
for the other states, gives
\begin{equation}
\sum_{B,\,(-1)^B=(-1)^N}
  \left(\begin{array}{c}
B+2n-1\\
B\end{array}\right)
\left(\begin{array}{c}
(N-B)/2+2n\\
2n
\end{array}\right).
\label{permdeg}
\end{equation}              
The sum over $B$ satisfying $(-1)^B=(-1)^N$ diverges as $N\rightarrow\infty$. 
Note that for $n\geq 1$, it is possible to break all the pairs and make all the
spins polarized, still with zero energy. In other words, the 
quasihole states of the spin-polarized $\nu=1$ state form a subset of the 
zero-energy states for the three-body Hamiltonian. 

As we have seen in the main text, the edge states are closely related to the
bulk quasihole states. For the permanent state, we will just state that there
are $4q$ sectors of edge states involving unpaired bosons, much like the other
examples in Ref.\ \cite{milr}. In the twisted sector, there are zero modes
which can be occupied with arbitrarily many bosons of either spin. Turning to
the ground states on the torus, there are again $4q$ sectors for $N$ even. The
ground states for ${\bf k}\neq0$ are an obvious generalization of those for the
$331$ state, containing a permanent instead of a determinant. These states are
again spin singlets. For ${\bf k}=0$, the construction that worked for 331 made
essential use of antisymmetrization, and does not work here. Instead, the only
possibility is to break all the pairs and put the bosons in the constant
single-particle state on the torus. The spin state is then totally symmetric,
so we obtain a spin $N/2$ multiplet of states. This construction also
works for $N$ odd.  

The large degeneracies of states in certain sectors in this system make sense
in the interpretation in terms of correlators in CFT. The theory relevant to
the permanent is a $\beta$-$\gamma$ ghost system \cite{mr}, where $\beta$ 
and $\gamma$ are free bosonic fields of conformal weight $1/2$, so that this 
theory is nonunitary. It realises Kac-Moody symmetry at level $k=-1/2$
\cite{mr} (not to
be confused with the vector ${\bf k}$ on the torus). The latter
theory includes the ``spin'' fields for the $\beta$-$\gamma$ system, which
behave as infinite-dimensional fractional-spin representations of SU(1,1) 
symmetry, that is related to, but not quite the same as, the SU(2) symmetry 
in which we are interested. Consequently the infinite degeneracies when 
quasiholes are present, or in the ground state on the torus in the 
corresponding sector, are not surprising. 

Another interesting question is the excitation spectrum
when the zero-energy states of our Hamiltonian are fully (or even just
macroscopically) spin-polarized, as occurs in the presence of 2 quasiholes or 
on the torus. Since the ground state breaks the spin-rotation symmetry, and the
Hamiltonian is short-range, we expect low-energy spin-wave excitations to
exist, in which one or more spins are flipped. For a generic Hamiltonian, 
one would expect these to occur at low but nonzero energy, with a gapless 
quadratic dispersion relation as in a quantum ferromagnet. Then the system 
would not be fully gapped for spin excitations, unlike (we believe) the 
other systems studied in this paper. It is interesting to note the relation 
with the $\nu=1$ spin-polarized system that has been much studied recently 
\cite{sondhi}. For a two-body Hamiltonian consisting of a projection operator 
onto zero relative angular momentum for two particles of total spin 0, the 
spin-polarized filled Landau level state, and quasihole states including 
reversed spins (skyrmions) are zero-energy eigenstates (note that the number 
of quasiholes for this system is identified with the number of flux added 
to the polarised ground state, which is smaller by 1 than the number of flux 
added to the permanent ground state). For each number of added flux, these 
are exactly the same as the states above with no unbroken pairs. There are 
also the expected spin-wave excitations at nonzero energy. However, in the 
case of the quasihole states of the Hamiltonian we have studied here, we 
already know that there are other states with some additional reversed spins 
at {\em zero} energy; they are simply the states where not all the pairs are 
broken. Clearly, these states would not be zero-energy states for the two-body 
Hamiltonian. It is tempting to identify them with a subset of the spin-wave 
states. 

{}From the summand in (\ref{permdeg}), we can obtain the number of zero-energy 
states for each 
number of unbroken pairs, $(N-B)/2$. It is instructive to begin with the case 
$n=1$ that corresponds to the sector containing the fully-polarized filled 
Landau level state. As we increase $(N-B)/2$ {}from zero, we expect that 
the total spin must decrease due to the formation of singlet pairs. The 
first binomial coefficient in eq.\ (\ref{permdeg}) is equal to $B+1$ which 
is the degeneracy of a single multiplet of spin $S=B/2$. The second binomial 
coefficient in eq.\ (\ref{permdeg}) is the orbital degeneracy of the 
quasiholes of the permanent, which we here interpret as the number of ways of
placing $(N-B)/2$ bosons in $2n+1=3$ orbitals. The bosons are to be viewed as
the spin waves. The three orbitals form an $L=1$ multiplet. The spin wave
excitations in general could have angular momenta $0$, $1$, \ldots; the $L=0$
ones simply represent a global rotation of the spin and have already been
counted in the degeneracy of each spin multiplet. We conclude that, for our
Hamiltonian, $L=1$ spin waves have zero energy at $n=1$. The other spin wave
states would have to be obtained by a collective excitation of the condensate of
singlet pairs, or spin waves (depending on our point of view), which excites
one or more of them to higher angular momentum. A similar picture holds for
$n>1$. Based on these arguments, we do expect a gapless branch of low-energy 
spin waves to be present in the spectrum of elementary excitations of the 
spin-polarized zero-energy states for our Hamiltonian, on the sphere, and 
also on the torus. 

The three-body Hamiltonian (\ref{perm3bodH}) above has been diagonalized 
numerically for $q=1$ and $N$ up to 12, with $V=1/18$, for various numbers 
of quasiholes; 
results for $n=0$ and $n=1$ are shown in Figures \ref{fig:perm0qh} and 
\ref{fig:perm2qh}. For $n=1$, the degeneracies of the zero-energy 
quasihole states have been confirmed, and in addition (see Fig.\
\ref{fig:perm2qh}) there are
low-lying states, so the system is not obviously gapped. In fact, by examining
the states at $n=1$ with $S_z=N/2-1$, that is one less than the maximum value,
which we expect to be the single-spin-wave states of the ferromagnet, we obtain
a dispersion relation of the spin waves, which is shown as the lower-right 
inset in Fig.\ \ref{fig:perm2qh}. It has the expected form of a
finite-size version of a gapless branch of states, and has zero energy
for both $L=0$ and $1$; the latter property may mean that the dispersion
relation has vanishing coefficient of wavevector-squared in the thermodynamic
limit (this coefficient is proportional to the spin stiffness in the
ferromagnet). We note that these states penetrate quite far into the full
spectrum. The apparent slight gap above the zero-energy states for $L\neq 5$ 
is in fact just
a finite size effect, since at least the $S_z=N/2-1$ states must form a 
gapless branch as $N\rightarrow\infty$. The low-lying states in the full 
spectrum should be other multi-spin-wave states.

For the ground state sector of the permanent, there is a nondegenerate
zero-energy state, as in the other cases studied in this paper, but there is
also an apparent gapless branch of states at low $L$ (see Fig.\
\ref{fig:perm0qh}). {}From the
point of view of the ferromagnet, this $N_\phi$ value represents a
quasielectron state, which would be an antiskyrmion. For the two-body
Hamiltonian, the antiskyrmions form a set of states with $L=S$, but which do
not have zero energy or exactly soluble wave functions. The lowest-energy 
states in Fig.\ \ref{fig:perm0qh} have $L=S$ for $L=0$, $1$, $2$, $3$, but 
$S=1$ for $L=4$, 
though here an $S=4$ state lies at slightly higher energy. Moreover, the $L=0$ 
ground state has overlap-squared $0.81$ with that for the Coulomb interaction 
at the same $N$, $N_\phi$. There is thus some evidence that this branch of 
states represents something similar to an antiskyrmion.

The proximity to the ferromagnet, albeit at a different number of flux
($N_\phi=q(N-1)$ for the ferromagnet), suggests that the system is 
at the transition to the ferromagnet. Indeed, if the two-body Hamiltonian
containing $V_0$ only (for $q=1$, or the usual generalizations for other $q$) 
is added to the three-body, then for $V_0>0$
the ferromagnet and its quasihole excitations (skyrmions) will be the only
remaining zero-energy states, and the $L=1$ spin waves (at $N_\phi=q(N-1)$)
will have non-zero energy. On the torus, the $4q$-fold degenerate sectors 
will be split to leave only the $q$ states with ${\bf k}=0$, which have 
maximum spin $N/2$.
For $V_0$ negative, we expect the splittings to reverse sign, and the ground
states on the torus, or on the sphere with quasiholes will presumably be
unpolarized; in this region no exact wave functions are available for the
lowest energy eigenstates. In other words, the polarized and unpolarized states
will differ in energy density. We expect that the unpolarized states are
paired, and indeed the attractive pseudopotential $V_0<0$ should favour the
pairs. In view of the higher energy density of the polarized ground state on the
torus, we guess that these states are no longer in the same universality class
as the permanent state, but may be a simple Halperin-type state, a Laughlin
state of charge-2 boson pairs. This probably would occur because the attractive
potential decreases the size of the pairs, and modifies the pairing function
{}from the simple form $(z_i^\uparrow-z_j^\downarrow)^{-1}$ found hitherto. As
nonabelian statistics probably relies upon the long-range character of this
part of the wave function, it could disappear under this perturbation. In any
case, the ferromagnetic order parameter is constant for $V_0>0$, and will
vanish for $V_0<0$, which indicates a first-order phase transition.

The transition also has a simple interpretation in terms of composite bosons.
In the permanent state, the bosons are paired, but when one flux quantum is
added, there are broken pair states of the same energy. In these states, the
unpaired composite bosons occupy the $L=0$ zero mode, and can be viewed as a
Bose condensate. Since they carry spin 1/2, such a condensate is necessarily a 
ferromagnet, and when more flux are added, the skyrmion zero-energy states are
obtained. Thus the permanent three-body Hamiltonian can have pairs but the
bosons can also unpair and form a condensate. In Bose liquid systems, the
condensation of single bosons is the usual occurence. It seems that the 
permanent state is on the borderline between a single-particle Bose condensate
and a condensate of pairs only. It is possible that, while the ferromagnetic
order indicates a first order transition, the behavior of the pair order
parameter (specifically, the size of the pairs), or of some properties 
on the ferromagnetic side, possibly related to a spin stiffness going to zero, 
could be characteristic of a second-order transition, with
a length that diverges at the transition point.



\begin{figure}
\caption{Two-particle correlation function $g(r)$ for the Pfaffian state, with
$N_\phi=2(N-1)-1$ (i.e.\ $\nu=1/2$), for $N=10$ (dashed line), $12$ 
(dot-dashed line), $14$
(solid line), versus great circle distance on the sphere. The curves for $N=12$
and $14$ are almost indistinguishable.}
\label{fig:pfg(r)}
\end{figure}

\begin{figure}
\caption{Spectrum of the three-body Hamiltonian for the Pfaffian state of
fermions with $\nu=1/2$, for $N=12$, $N_\phi=21$, that is, no quasiholes. 
The inset enlarges the low-lying levels.}
\label{fig:pf0qh}
\end{figure}

\begin{figure}
\caption{As in Fig.\ \protect\ref{fig:pf0qh}, but with $N=10$, $N_\phi=19$, 
that is, $4$ quasiholes.}
\label{fig:pf4qh}
\end{figure}

\begin{figure}
\caption{Low-lying excited states of the three-body Hamiltonian 
for the Pfaffian ground state at $\nu=1/2$
($N_\phi=2(N-1)-1$) for $N=10$ and $12$, plotted against $k=L/R$.}
\label{fig:pfneutsize}
\end{figure}

\begin{figure}
\caption{As in Fig.\ \protect\ref{fig:pfneutsize}, but for the energy 
$\Delta$ of the lowest-energy excited state versus $1/N$.}
\label{fig:pfneutgapscal}
\end{figure}

\begin{figure}
\caption{Ground state energy of the three-body Hamiltonian for the Pfaffian
state at $N_\phi=2(N-1)-2$, that is, $\nu=1/2$ with two quasielectrons added,
plotted against $1/N$.}
\label{fig:pfqelgapscal}
\end{figure}

\begin{figure}
\caption{Spectrum of the hollow-core model for the HR state of fermions 
with $\nu=1/2$, for $N=8$, $N_\phi=12$, that is, no quasiholes. 
The inset enlarges the low-lying levels.}
\label{fig:hr0qh}
\end{figure}

\begin{figure}
\caption{As in Fig.\ \protect\ref{fig:hr0qh}, but with
$N=6$, $N_\phi=12$, that is, $8$ quasiholes.}
\label{fig:hr8qh}
\end{figure}

\begin{figure}
\caption{Spectrum of the Ho model, for four values of $\theta$ 
which parametrizes the $\bf d$ vector, all for $N=6$, $N_\phi=9$, that is 
$\nu=1/2$ and no quasiholes.}
\label{fig:ho0qh}
\end{figure}

\begin{figure}
\caption{Same as Fig.\ \protect\ref{fig:ho0qh}, but enlarged to show low-lying 
levels.}
\label{fig:ho0qhenl}
\end{figure}

\begin{figure}
\caption{Spectrum of the Ho model plus the three-body interaction 
at the Pfaffian point $\theta=\pi/4$, again for $N=6$, $N_\phi=9$. 
The inset enlarges the low-lying levels.}
\label{fig:ho+3bod0qh}
\end{figure}

\begin{figure}
\caption{As in Fig.\ \protect\ref{fig:ho+3bod0qh}, but $N=6$, $N_\phi=11$, 
that is, $4$ quasiholes.}
\label{fig:ho+3bod4qh}
\end{figure}

\begin{figure}
\caption{Spectrum of the three-body interaction for the permanent state of 
fermions with $\nu=1$, for $N=12$, $N_\phi=10$, that is, no quasiholes. 
The inset enlarges the low-lying levels.}
\label{fig:perm0qh}
\end{figure}

\begin{figure}
\caption{As in Fig.\ \protect\ref{fig:perm0qh}, but with $N=10$, $N_\phi=9$, 
that is, $2$ quasiholes. Upper-left inset: low-lying levels. Lower-right inset
shows only the states with $S_z=N/2-1$, i.e.\ single spin flips (or spin waves) 
of the fully-polarized state.}
\label{fig:perm2qh}
\end{figure}


\begin{table}
\caption{Numbers of multiplets of states of total angular momentum $L$ 
at zero energy for the three-body Hamiltonian for the $q=2$ Pfaffian state on
the sphere, for $N_\phi=2(N-1)-1+n$, i.e.\ $2n$ quasiholes.}
\label{tab:pfzeroen}
\end{table}

\begin{table}
\caption{Numbers of multiplets of states of total angular momentum $L$ 
and total spin $S$ at zero energy for the hollow-core Hamiltonian for the $q=2$
HR state on the sphere, for $N_\phi=2(N-1)-2+n$, i.e.\ $2n$ quasiholes.}
\label{tab:hrzeroen}
\end{table}

\end{document}